\documentclass[12pt,english,longbibliography,nofootinbib,superscriptaddress,12pt,sort&compress,showkeys]{article}
\usepackage{ae,aecompl}
\usepackage[T1]{fontenc}
\usepackage[latin9]{inputenc}
\usepackage[active]{srcltx}
\usepackage{amsmath}
\usepackage{graphicx}
\usepackage{esint}
\usepackage[numbers]{natbib}

\makeatletter

\providecommand{\tabularnewline}{\\}
\newcommand{\lyxdot}{.}

\newcommand{\lyxaddress}[1]{
	\par {\raggedright #1
	\vspace{1.4em}
	\noindent\par}
}

\usepackage{aecompl}\usepackage{epstopdf}

\usepackage[raggedrightboxes]{ragged2e}

\usepackage{babel}

\makeatother

\usepackage{babel}
\begin{document}
\title{A model of thermodynamic stabilization of nanocrystalline grain boundaries
in alloy systems}
\author{Omar Hussein and Yuri Mishin}
\maketitle

\lyxaddress{Department of Physics and Astronomy, MSN 3F3, George Mason University,
Fairfax, Virginia 22030, USA}
\begin{abstract}
\noindent Nanocrystalline (NC) materials are intrinsically unstable
against grain growth. Significant research efforts have been dedicated
to suppressing the grain growth by solute segregation, including the
pursuit of a special NC structure that minimizes the total free energy
and completely eliminates the driving force for grain growth. This
fully stabilized state has been predicted theoretically and by simulations
but is yet to be confirmed experimentally. To better understand the
nature of the full stabilization, we propose a simple two-dimensional
model capturing the coupled processes of grain boundary (GB) migration
and solute diffusion. Kinetic Monte Carlo simulations based on this
model reproduce the fully stabilized polycrystalline state and link
it to the condition of zero GB free energy. The simulations demonstrate
the emergence of a fully stabilized state by the divergence of capillary
wave amplitudes on planar GBs and by fragmentation of a large grain
into a stable ensemble of smaller grains. The role of solute diffusion
in the full stabilization is examined. Possible extensions of the
model are discussed. 
\end{abstract}

\section{Introduction\label{sec:Intro}}

Many nanocrystalline (NC) materials exhibit physical and mechanical
properties superior to those of coarse-grained materials. However,
a major obstacle to broader applications of NC materials is their
instability against grain growth at elevated temperatures. The excess
grain boundary (GB) free energy $\gamma$ gives rise to capillary
forces driving GB migration to reduce the GB area. The excess free
energy in NC materials is extremely high, often making the grain growth
virtually unstoppable. Significant research efforts have been dedicated
to suppressing the grain growth by alloying the material with suitable
solutes. Such solutes segregate to GBs and either lower their free
energy or reduce their mobility (or both). Grain growth retardation
by alloying has been achieved in numerous alloy systems. Unfortunately,
the stabilization by alloying becomes less effective at high temperatures
because GB segregation weakens while GB mobility increases. Another
limiting factor is the competition between GB segregation and bulk
phase transformations.

Three decades ago, Weissmuller \citep{Weissmuller1994,Weissmuller:1993aa}
suggested that, under certain conditions, the total free energy of
a NC alloy can reach a minimum at a \emph{finite} grain size. In this
special state, the driving force for grain coarsening vanishes, and
the material becomes thermodynamically stable against grain growth.
We refer to this special state as \emph{fully stabilized}. Thermodynamic
analysis \citep{Kirchheim:2002aa,Liu:2004aa,Kirchheim2007a,Kirchheim2007b,Krill:2005aa,Schvindlerman2006,Detor2007,Chookajorn2012,Chookajorn2014,Kalidindi:2017aa,Kalidindi:2017bb,Kalidindi:2017cc,Murdoch:2013aa,Trelewicz2009,Perrin-2021}
confirmed the theoretical possibility of full stabilization. The conditions
of full stabilization were investigated in great detail within the
regular solution model \citep{Detor2007,Chookajorn2012,Chookajorn2014,Kalidindi:2017aa,Kalidindi:2017bb,Kalidindi:2017cc,Murdoch:2013aa,Trelewicz2009,Perrin-2021,Koch2008,Darling2014,Saber2013,Saber2013a,Xing:2018aa,Zhou:2014aa},
and the results were presented as stability diagrams convenient for
alloys design. Metropolis Monte Carlo simulations \citep{Chookajorn2014,Trelewicz2009,Kalidindi:2015aa}
have demonstrated that, under certain combinations of the regular
solution parameters, a polycrystal can reach equilibrium with a finite
grain size, confirming the full stabilization. On the other hand,
despite the abundant experimental evidence of grain growth inhibition
by alloying, there is no convincing evidence that a fully stabilized
NC state has been realized. Further theoretical and simulation efforts
are required to better understand the nature of the full stabilization
and assess the feasibility of its practical implementation.

The goal of this paper is to investigate two aspect of the NC stabilization
problem. The first one is related to the conditions of full stabilization.
Assuming that GB segregation is uniform across the entire GB area
in the sample (which is a strong approximation), it can be shown (see
Appendix) that reaching a total free energy minimum is equivalent
to the condition of zero GB free energy (per unit area): $\gamma=0$.
To create this state, the system must be closed (no solute exchange
with the environment) and have a nano-scale grain size, so that the
solute storage capacities of the GBs and the grains would be comparable.
In an open system, an initially planar GB must develop diverging capillary
fluctuations as $\gamma$ approaches zero. In a closed system, the
capillary fluctuations must increase but remain finite. Once the fully
stabilized state is reached, the $\gamma=0$ condition can be verified
by directly computing the GB free energy (e.g., by thermodynamic integration).
To our knowledge, this type of analysis has not been carried out. 

The second aspect is related to the role of solute diffusion in the
NC stability. The previous theoretical analyses \citep{Detor2007,Chookajorn2012,Chookajorn2014,Kalidindi:2017aa,Kalidindi:2017bb,Kalidindi:2017cc,Murdoch:2013aa,Trelewicz2009,Perrin-2021,Koch2008,Darling2014,Saber2013,Saber2013a,Xing:2018aa,Zhou:2014aa,Kirchheim2007a,Kirchheim2007b,Kirchheim:2002aa,Liu:2004aa,Schvindlerman2006,Weissmuller1994,Weissmuller:1993aa,Krill:2005aa}
had a \emph{global} character. The free energy was minimized with
respect to the total GB area (assumed to be proportional to the average
grain size) under the constraints of GB-solute equilibrium and a fixed
total amount of the solute. This constrained minimization tacitly
assumes that local GB displacements are accompanied by instant redistribution
of the solute across the entire system, quickly creating a new GB-solute
equilibrium corresponding to the changed GB area. The Monte Carlo
simulations \citep{Chookajorn2014,Trelewicz2009,Kalidindi:2015aa}
are similarly global: the solute and solvent atoms are randomly exchanged
across the entire system, ensuring fast global re-equilibration in
response to variations in the number of GB sites. The assumption of
instant re-equilibration is incompatible with the finite rate of solute
diffusion. The GB displacements are \emph{local} events accompanied
by absorption or rejection of some amount of the solute. Diffusion-controlled
redistribution of this solute cannot be fast enough to immediately
restore the GB-solute equilibrium everywhere in the sample. Equilibrium
will be lost, at least locally and temporarily. It is unclear \emph{a
priori} whether the structure minimizing the total free energy with
respect to variations in the total GB area will be stable against
other variations, such as local perturbations of the grain size distribution.
Answering this question requires a model explicitly treating both
solute diffusion and GB migration.

In this paper, we develop a simple model of a polycrystalline alloy
that captures solute diffusion and solute interactions with stationary
and moving GBs. The model is more conceptual than quantitative. It
is only intended to probe the fundamental aspects of the full GB stabilization
mentioned above. It combines a two-dimensional (2D) Ising model on
a square lattice with a solid solution model. Previous Monte Carlo
simulations using the Ising or Potts models with solutes \citep{Mendelev:2001aa,Mendelev:2001wk,Trelewicz2009,Chookajorn2014,Liu:1998ab,Liu:1999aa}
employed the Metropolis algorithm. The latter is appropriate for equilibrium
systems but does not predict the correct dynamics away from equilibrium.
In particular, the Metropolis algorithm does not correctly describe
diffusion in structurally inhomogeneous systems. To address this limitation,
we develop a kinetic Monte Carlo (KMC) model describing concurrent
operation of GB migration and solute diffusion. We apply this model
to calculate the thermodynamic properties of alloy GBs, reproduce
the fully stabilized polycrystalline state, verify the link between
this state and the $\gamma=0$ condition, and examine the role of
solute diffusion in the full stabilization.

\section{Model formulation}

\subsection{Structure and thermodynamics of the system}

We consider a 2D array of square cells of size $a$ arranged in a
square pattern. The array is $X$ cells long in the direction parallel
to the Cartesian $x$-axis and $Y$ cells long parallel to the Cartesian
$y$-axis. Thus, the physical dimensions of the array are $aX\times aY$.
The cells are enumerated by a single index $k=1,2,..,N$, where $N=XY$
is the total number of cells. Periodic boundary conditions are applied
to mimic an infinitely large system. 

The cells are interpreted as small crystalline blocks with two possible
lattice orientations. We refer to these orientations as black and
white and label them with a ``spin'' variable $\sigma_{k}$ taking
the values of $\sigma_{k}=1$ and $\sigma_{k}=-1$ for the black and
white orientations, respectively. The average orientation (color)
of the system is characterized by the variable
\begin{equation}
I=\dfrac{\sum_{k}\sigma_{k}}{N},\label{eq:0}
\end{equation}
with $I=1$ and $I=-1$ for the all-black and all-white single-crystalline
states, respectively. Eq.(\ref{eq:0}) can also calculate the average
color of a subset of cells.

In addition to the ``spin'' variable $\sigma_{k}$, we introduce another
variable $n_{k}$ defined as the number of unlike (opposite color)
nearest neighbors of the cell $k$. The $n_{k}$ values are related
to the $\sigma_{k}$ values by
\begin{equation}
n_{k}=2-\dfrac{\sigma_{k}}{2}\sum_{(kl)}\sigma_{l},\label{eq:2}
\end{equation}
where the symbol $(kl)$ indicates summation over the four neighbors
$l$ of the cell $k$. If all neighbors have the same orientation
as the cell $k$, then $n_{k}=0$; if all neighbors have the opposite
orientation, then $n_{k}=4$. All other cases lie in between. 

Next, we assume that unlike nearest-neighbor cells interact with a
repulsive potential driving the system to a color separation. Specifically,
each pair of nearest-neighbor cells with opposite orientation interacts
with a positive energy $2J_{gg}>0$, while nearest-neighbor pairs
of the same orientation do not interact. Thus, for a given distribution
of black and white cells, the crystallographic energy of the system
is 
\begin{equation}
E_{\mathrm{cryst}}=\sum_{k}J_{gg}n_{k}.\label{eq:1}
\end{equation}
Note that $E_{\mathrm{cryst}}=0$ in a single-crystalline state and
reaches the maximum value of $E_{\mathrm{cryst}}=4J_{gg}N$ for a
checkerboard arrangement of the cells. 

This model is isomorphous to the 2D Ising model on a simple square
lattice, in which the total energy is 
\begin{equation}
E_{\mathrm{Ising}}=-\dfrac{1}{2}\sum_{k,l}J\sigma_{k}\sigma_{l}-\sum_{k}h\sigma_{k}.\label{eq:1-1}
\end{equation}
The interaction parameter $J$ and the external field $h$ are related
to our model by $J=J_{gg}$ and $h=0$. The average orientation $I$
in Eq.(\ref{eq:0}) is similar to the magnetization in the Ising model.

\begin{figure}
\begin{centering}
\includegraphics[width=0.9\textwidth]{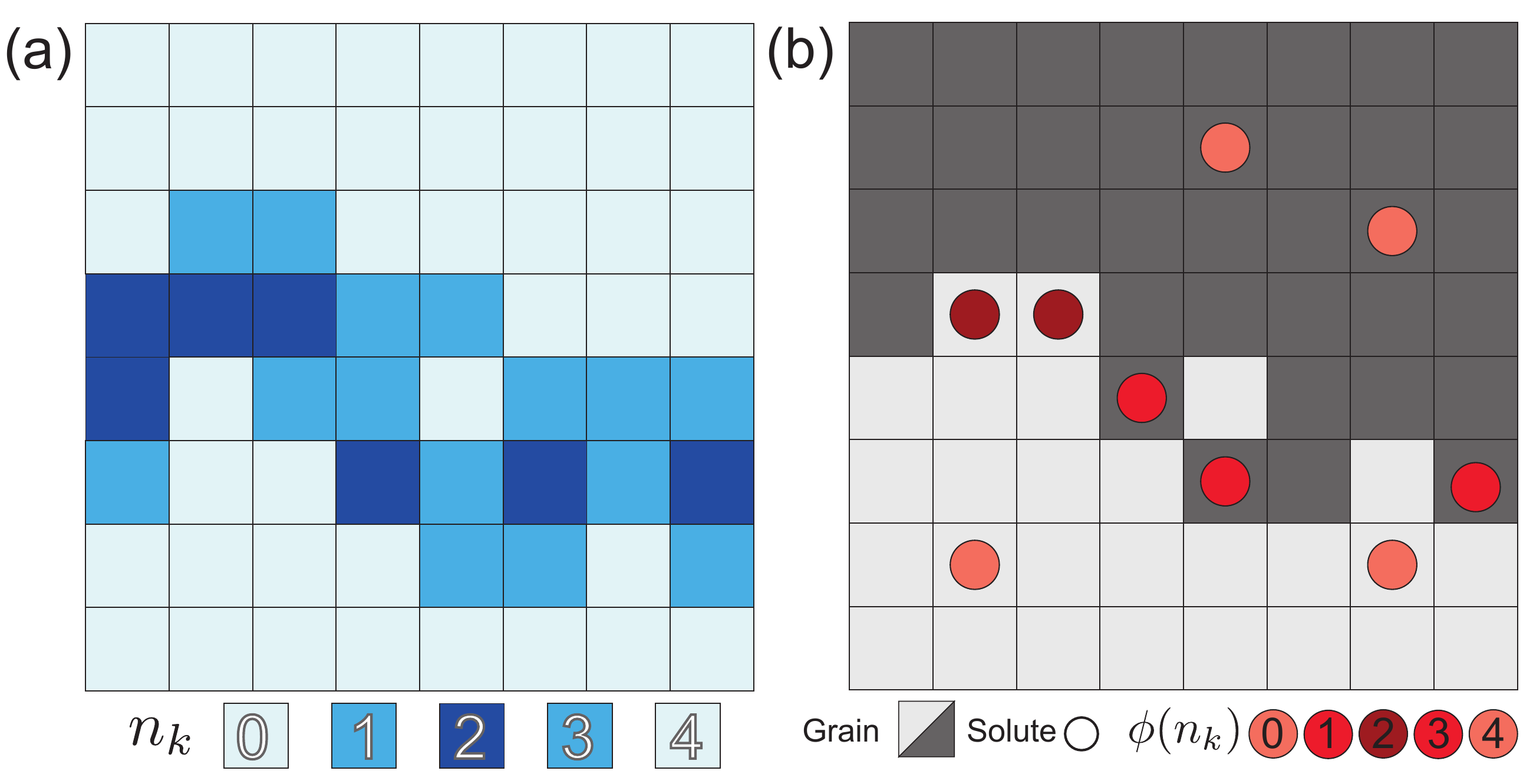}
\par\end{centering}
\caption{Schematic presentation of a typic grain boundary in the present model.
(a) Color-coded values of the parameter $n_{k}$ showing the concentration
of cells with $n_{k}=2\pm1$ in the GB region. (b) Typical distribution
of solute atoms color-coded by the parameter $\phi(n_{k})$, demonstrating
solute attraction to the GB region. \label{fig:schematic-GB}}
\end{figure}

If the system is partitioned into black and white domains, we interpret
them as grains. Accordingly, boundaries between such domains, on which
the crystallographic orientation switches from black to white, represent
GBs. To track the motion of GBs, we introduce the GB locator function
\begin{equation}
\phi(n_{k})=1-\frac{1}{4}\left(n_{k}-2\right)^{2}\label{eq:phi}
\end{equation}
representing the likelihood that a given cell $k$ lies in a GB region.
This function is zero inside a single crystal ($n_{k}=0$) and for
an isolated cell of opposite color relative to its environment ($n_{k}=4$).
Both cases are representative of a grain interior. The function reaches
the maximum value of $1$ for cells with $n_{k}=2$. Such cells are
transitional between the single-crystalline states on either side
of the GB and can be associated with the GB region. Fig.~\ref{fig:schematic-GB}(a)
depicts a portion of a typical GB with the cells colored according
to the $n_{k}$ value. Note that the cells with $n_{k}=2\pm1$ decorate
the GB region, permitting its automated detection and tracking during
simulations.

We next extend the model to include solute atoms, which can occupy
some of the cells with one atom per cell. If the system contains $N_{s}\leq N$
solute atoms distributed among the cells, the average solute concentration
is $c=N_{s}/N$. To model GB segregation, we assume that the GBs create
a potential field attracting the solute atoms. To represent this field,
the following term is added to the total energy:
\begin{equation}
E_{sg}=-\sum_{k}J_{sg}\xi_{k}\phi(n_{k}).\label{eq:solute-GB}
\end{equation}
Here, $\xi_{k}$ is a solute locator whose values are $\xi_{k}=1$
if the cell $k$ is occupied by a solute atom and $\xi_{k}=0$ otherwise.
Each term is multiplied by the GB locator $\phi(n_{k})$, so the product
$\xi_{k}\phi(n_{k})$ identifies the GB cells occupied by solute atoms.
The parameter $J_{sg}>0$ controls the strength of GB segregation.
Fig.~\ref{fig:schematic-GB}(b) illustrates the solute attraction
to the GB, with colors representing the values of $J_{sg}\xi_{k}\phi(n_{k})$.

The model also includes interaction among the solute atoms. Only nearest-neighbor
atoms are allowed to interact, and their interactions are assumed
to be repulsive, driving the system to phase separation below a critical
point. Furthermore, the solute-solute interactions generally differ
inside the grains and in GBs. Given that the GBs are associated with
$\phi(n_{k})$ values close to $2$, the strength of interactions
must correlate with the function $\phi(n_{k})$. Based on these considerations,
we postulate the following form of the total energy of solute-solute
interactions:
\begin{equation}
E_{ss}=\dfrac{1}{2}\sum_{k}\xi_{k}\sum_{(kl)}\xi_{l}J_{ss}\left[1+\eta\phi(n_{k,l})\right].\label{eq:s-s}
\end{equation}
In this equation, the solute locators $\xi_{k}$ and $\xi_{l}$ select
nearest neighbor cells occupied by solute atoms, the factor of $1/2$
eliminates the double-count of the atomic pairs, and we have introduced
\begin{equation}
n_{k,l}=\frac{n_{k}+n_{l}}{2}\label{eq:nkl}
\end{equation}
as the average number of black-white bonds of the two solute atoms.
The parameter $J_{ss}>0$ is the strength of solute-solute interactions
when both atoms are inside a grain, in which case $\phi(n_{k,l})=0$
and the term in the square brackets equals unity. If one or both atoms
are in or near a GB, $\phi(n_{k,l})$ is positive and the interaction
between the two atoms is modified, becoming stronger or weaker, depending
on whether $\eta$ is chosen positive or negative. If $n_{k}=n_{l}=2$,
the interaction parameter becomes $J_{ss}\left[1+\eta\right]$. This
setup allows us to control the solution thermodynamics in the grains
and in the GB segregation atmosphere. 

To summarize, the model describes a non-ideal solid solution in a
polycrystalline structure with solute-GB interactions. Thermodynamic
properties of the solution are different in the GBs and inside the
grains. Each microstate of the system is defined by the set of parameters
$\{\sigma_{k},\xi_{k}\}$. The total energy of the system in a given
microstate is
\begin{eqnarray}
E & = & E_{\mathrm{cryst}}+E_{sg}+E_{ss}\label{eq:E_tot}\\
 & = & \sum_{k}J_{gg}n_{k}-\sum_{k}J_{sg}\xi_{k}\phi(n_{k})+\dfrac{1}{2}\sum_{k}\xi_{k}\sum_{(kl)}\xi_{l}J_{ss}\left[1+\eta\phi(n_{k,l})\right].\nonumber 
\end{eqnarray}
The function $\phi(n_{k})$ is defined by Eq.(\ref{eq:phi}) and the
numbers $n_{k}$ and $n_{k,l}$ are given by Eqs.(\ref{eq:2}) and
(\ref{eq:nkl}), respectively. The input material parameters of the
model are $J_{gg}$, $J_{sg}$, $J_{ss}$, and $\eta$, which characterize
the orientational interactions, the GB segregation energy, the solute-solute
binding, and the GB thermodynamics, respectively.

\subsection{Time evolution of the system}

To model the dynamics of our system, we couple it to a thermostat
at a temperature $T$ and let it evolve by overcoming energy barriers
between microstates by thermal fluctuations. Only transitions between
microstates separated by a single energy maximum are considered. Furthermore,
only two types of transition are allowed, which we call a flip and
a jump. In a flip, a cell switches its color to the opposite one.
Sequences of flips from black to white, or vice versa, can result
in GB migration. In a jump event, a solute atom jumps to a nearest-neighbor
cell. Such jumps constitute a mechanism of solute diffusion.

\begin{figure}
\noindent \begin{centering}
\includegraphics[width=0.52\textwidth]{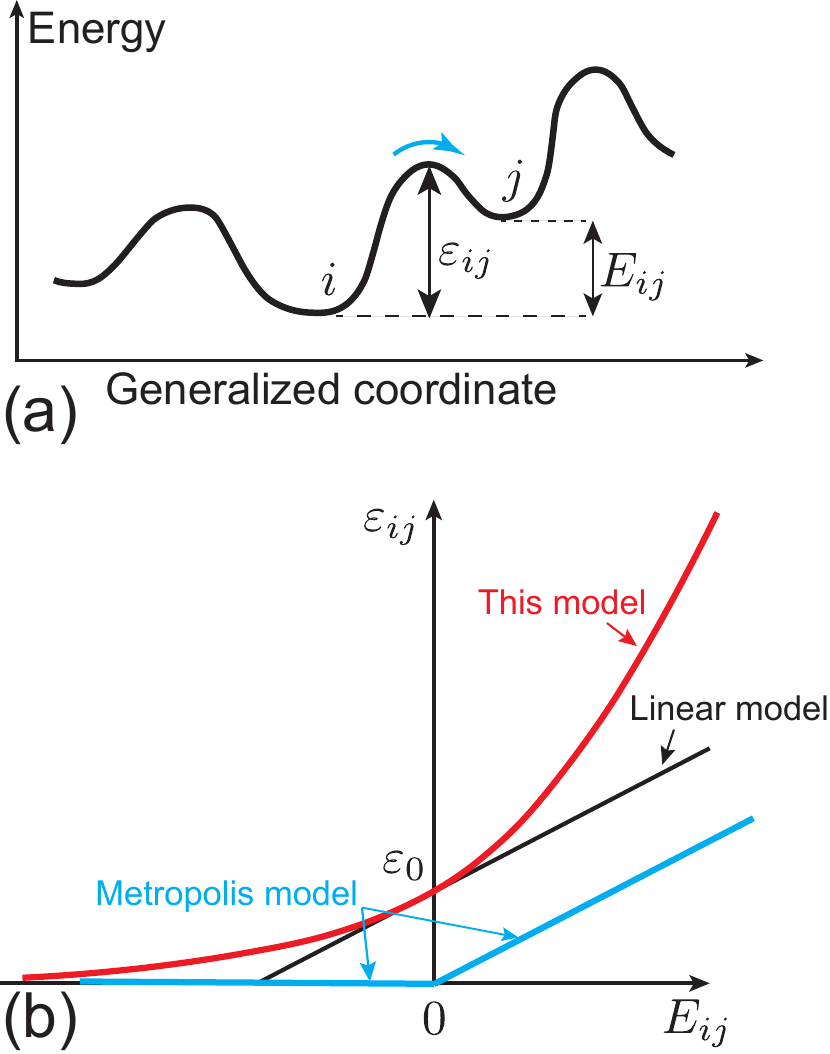}
\par\end{centering}
\caption{(a) Schematic 1D representation of energy landscape of a system jumping
between energy minima by thermal fluctuations. $\varepsilon_{ij}$
is the transition barrier from state $i$ to state $j$ with energies
$E_{i}$ and $E_{j}$, respectively. (b) The transition barrier as
a function of energy difference $E_{ij}=E_{j}-E_{i}$ in the present
model compared with the linear model (Eq.(\ref{eq:linear-model}))
and the Metropolis model (Eq.(\ref{eq:p_ij_flip})).\label{fig:barrier-models}}

\end{figure}

We adopt the harmonic transition state theory \citep{Vineyard:1957vo},
by which the transition rate $\nu_{ij}$ from a microstate $i$ to
a microstate $j$ is

\begin{equation}
\nu_{ij}=\nu_{0}\exp\left(-\dfrac{\varepsilon_{ij}}{k_{B}T}\right),\label{eq:1.01-1}
\end{equation}
where $\varepsilon_{ij}$ is the transition barrier, $k_{B}$ is Boltzmann's
constant, and $\nu_{0}$ is the attempt frequency. The latter is assumed
to be the same for all transitions. The transition barrier $\varepsilon_{ij}$
depends on the energy difference, $E_{ij}=E_{j}-E_{i}$, between the
states $i$ and $j$. If the system transitions to a lower-energy
state ($E_{j}<E_{i}$), the barrier $\varepsilon_{ij}$ must be lower
than for the reverse transition (Fig.~\ref{fig:barrier-models}(a)).
If the energy difference between the states is small, the linear approximation
\begin{equation}
\varepsilon_{ij}=\varepsilon_{0}+E_{ij}/2\label{eq:linear-model}
\end{equation}
is often used, in which $\varepsilon_{0}$ is the unbiased energy
barrier (when $E_{j}=E_{i}$). The unbiased barriers are generally
different for the flips and jumps; we denote them $\varepsilon_{0}^{g}$
(flip) and $\varepsilon_{0}^{s}$ (jump). 

The linear approximation in Eq.(\ref{eq:linear-model}) is only valid
for weakly driven systems. This work adopts a more general, nonlinear
model proposed in \citep{Mishin:2023ab,Mishin:2023aa}. In this model,
\begin{equation}
\varepsilon_{ij}=\begin{cases}
\varepsilon_{0}\exp\left(\dfrac{E_{ij}}{2\varepsilon_{0}}\right), & E_{ij}\leq0,\\
E_{ij}+\varepsilon_{0}\exp\left(-\dfrac{E_{ij}}{2\varepsilon_{0}}\right), & E_{ij}>0.
\end{cases}\label{eq:1.02}
\end{equation}
If the energy difference $E_{ij}$ is positive and large ($E_{ij}\gg\varepsilon_{0}$),
the barrier is high and close to $E_{ij}$. If $E_{ij}$ is negative
and large in magnitude ($-E_{ij}\gg\varepsilon_{0}$), the barrier
is exponentially small. In other words, a strong driving force suppresses
the transition barrier but never makes it exactly zero. The barrier-energy
relation within this model is illustrated in Fig.~\ref{fig:barrier-models}(b).
Note that this relation guarantees the detailed balance of the forward
and backward transitions.

It is known that GB diffusion is much faster than lattice diffusion
and has a lower activation energy \citep{Kaur95,Mishin97e,Mishin99f}.
The effect is called ``short circuit'' diffusion and is observed in
most (if not all) materials. To capture this effect, we make the unbiased
barrier of solute diffusion different in GBs and inside the grains.
To this end, we make $\varepsilon_{0}^{s}$ a function of $\phi(n_{k,l})$.
Namely, for a solute jump $k\rightarrow l$, the unbiased barrier
is
\begin{equation}
\varepsilon_{0}^{s}=\varepsilon_{00}^{s}\left[1-q\phi(n_{k,l})\right],\label{eq:Solute-barrier}
\end{equation}
where $\varepsilon_{00}^{s}$ is the barrier in a single crystal (for
which $\phi(n_{k,l})=0$), and $q$ is a parameter. The latter is
adjusted to reduce the GB diffusion barrier. Indeed, in GB regions
$\phi(n_{k,l})\approx1$ and thus $\varepsilon_{0}^{s}\approx\varepsilon_{00}^{s}\left[1-q\right]$.
Since $\varepsilon_{0}^{s}$ cannot be negative, $q$ must be smaller
than $1$. In this work, we take $q=1/2$, which gives $\varepsilon_{0}^{s}\approx\varepsilon_{00}^{s}/2$
in reasonable agreement with experimental data \citep{Kaur95,Mishin97e,Mishin99f}.
Note that the environmental adjustment of the diffusion barriers only
applies to solute atom jumps. The unbiased barrier of the flips ($\varepsilon_{0}^{g}$)
remains constant.

\section{Computer implementation of the model}

In this section, we discuss the computer implementation of this model.
The computations were performed in normalized variables whose relations
to the physical variables are summarized in Table \ref{tab:variables}.
All energies are normalized by $J_{gg}$, all distances by $a$, and
the unit of time is the inverse attempt frequency, $\nu_{0}^{-1}$.
The normalized variables are denoted by the same symbols as the physical
variables. This should not cause any confusion because we will only
use the normalized variables from now on.

Two types of simulation were performed: simulations of equilibrium
thermodynamic states of the system and simulations of the system dynamics.
Accordingly, two different algorithms were used, as described below.

\subsection{Thermodynamic property calculations\label{subsec:Metropolis}}

Equilibrium thermodynamic properties were calculated by Metropolis
Monte Carlo simulations in the grand canonical ensemble. Starting
from an initial state of the system, the simulations first brought
the system to thermodynamic equilibrium, followed by a long production
run to average the properties of interest. 

Two types of trial moves were implemented: a flip and insertion or
removal of a solute atom. At each Monte Carlo step, one of the two
types of move was selected with equal probability (except for the
solute-free simulations when only flips were considered). For a flip,
the color of a randomly selected cell was reversed, and the energy
change $E_{ij}$ was calculated. The attempt was accepted with the
probability 
\begin{equation}
p_{ij}=\min\left(1,e^{-\frac{E_{ij}}{T}}\right).\label{eq:p_ij_flip}
\end{equation}
For the solute insertion/removal, the occupation number $\xi_{k}$
of a randomly selected cell $k$ was switched to the opposite, and
the attempt was accepted with the probability
\begin{equation}
p_{ij}=\min\left(1,e^{-\frac{E_{ij}\pm\mu}{T}}\right),\label{eq:p_ij_insert}
\end{equation}
where $\mu$ is a preset value of the chemical potential. The sign
in the exponent depends on whether the solute atom was removed ($+$)
or inserted ($-$). Note that the flip case can be formally considered
as insertion/removal of a black cell with $\mu=0$, so the statistical
ensemble is grand-canonical in both cases. For both types of trial
move, if $p_{ij}$ was found to be $<1$, the acceptance decision
was made by generating a random number $r\in(0,1]$. The move was
accepted if $p_{ij}\geq r$. 

The equilibrium values of properties of interest were obtained by
simple averaging over the visited microstates. We emphasize that the
simulation implements a single trajectory in the configuration space
$\{\sigma_{k},\xi_{k}\}$. To ensure ergodicity of the sampling, the
simulation must be repeated multiple times from random initial states.

\subsection{Simulation of system dynamics}

The system evolution was studied using KMC simulations. The system
was coupled to a thermostat and chemically isolated while allowing
non-conservative changes in the cell orientations. In other words,
the statistical ensemble was canonical for solute jumps and grand-canonical
with $\mu=0$ for cell orientations. This allowed us to model the
concurrent processes of solute diffusion and grain evolution. 

A rejection-free KMC algorithm was implemented, which was similar
to the $n$-way algorithm \citep{bortz1975new}. The simulations comprised
the following steps: 
\begin{enumerate}
\item Initialize the simulation: set the clock to zero ($t_{0}=0$), set
the iteration number to zero ($m=0$), and choose an initial microstate
$i=\{\sigma_{k},\xi_{k}\}$. 
\item Use Eqs.~(\ref{eq:1.01-1}) and (\ref{eq:1.02}) to calculate the
set of rates $\left\{ \nu_{ij}^{s}\right\} $ for all possible jumps
of all solute atoms. Note that jumps to occupied cells are excluded
from the calculation. 
\item Calculate the total solute jump rate from the microstate $i$,
\[
R_{i}^{s}=\sum_{j=1}^{M_{i}^{s}}\nu_{ij}^{s},
\]
and the commutative jump probabilities
\[
P_{n}^{s}=\dfrac{1}{R_{i}^{s}}\sum_{j=1}^{n}\nu_{ij}^{s},\enskip n=1,2,...,M_{i}^{s}.
\]
Here, $M_{i}^{s}$ is the total number of allowed solute jumps from
the microstate $i$.
\item Perform similar calculations for the total flip rate $R_{i}^{f}$
and the cumulative transition probabilities $P_{n}^{f}$ ($n=1,2,...,N$)
for all flips from the microstate $i$. The required flip rates $\left\{ \nu_{ij}^{f}\right\} $
are obtained by flipping each cell, computing the energy change $E_{ij}$,
and using Eqs.~(\ref{eq:1.01-1}) and (\ref{eq:1.02}).
\item Find the expected times to escape from the microstate $i$ by making
a solute jump ($\tau_{i}^{s}$) and by a flip ($\tau_{i}^{f}$). To
this end, generate two random numbers $r_{1},r_{2}\in(0,1]$ and calculate
\[
\tau_{i}^{s}=-\dfrac{1}{R_{i}^{s}}\ln r_{1},\enskip\enskip\tau_{i}^{f}=-\dfrac{1}{R_{i}^{f}}\ln r_{2}.
\]
\item If $\tau_{i}^{s}<\tau_{i}^{f}$, implement a solute jump. To choose
which atom will jump, draw a third random number $r_{3}\in(0,1]$
and find index $n$ such that $P_{n-1}^{s}<r_{3}\leq P_{n}^{s}$ (assuming
$P_{0}^{s}=0$). Execute the solute jump $n$ by updating the respective
occupation numbers. If $\tau_{i}^{s}\geq\tau_{i}^{f}$, implement
a flip of the cell $n$ such that $P_{n-1}^{f}<r_{3}\leq P_{n}^{f}$
(assuming $P_{0}^{f}=0$). Execute the flip by updating $\sigma_{n}$
and recomputing the $n_{k}$ and $n_{k,l}$ values for the cells affected
by the flip.
\item Update the iteration number to $m+1$.
\item Advance the clock: $t_{m}=t_{m-1}+\tau_{m}$, where $\tau_{m}=\min(\tau_{i}^{s},\tau_{i}^{f})$.
\item Rename the new microstate to $i$.
\item Return to step 3. 
\end{enumerate}
We emphasize that the KMC simulations are more suitable for modeling
solute diffusion and GB migration than the Metropolis algorithm. The
latter correctly describes the system in thermodynamic equilibrium
and drives a non-equilibrium system towards equilibrium. However,
 the path towards the equilibrium state is not guaranteed to be physically
meaningful. When making the accept/reject decisions, the Metropolis
algorithm only considers the energies of the microstates and disregards
the unbiased transition barriers. Specifically, the transition barrier
to a higher-energy microstate is assumed to be equal to the energy
difference between the microstates, while transitions to lower-energy
microstates are barrier-free. This behavior is shown schematically
in Fig.~\ref{fig:barrier-models}(b). In reality, the transition
probabilities from a given microstate depend not only on the energies
of the available destination microstates but also on the barriers
separating those microstates from the current one. For example, if
two destination microstates have the same energy but one is separated
from the current microstate by a lower barrier, transition to the
latter microstate is more probable. The higher probability of lower-barrier
states with the same energy is built into the KMC algorithm proposed
here. By contrast, the Metropolis algorithm will treat both transitions
as equally probable. Due to this difference, the chains of transitions
generated by the two algorithms are generally different.

The KMC algorithm can also be applied to sample equilibrium states
and should give the same results as the Metropolis algorithm. Once
an equilibrium state is reached, the expected value of any property
$\mathcal{P}\left(\left\{ \sigma_{k},\xi_{k}\right\} \right)$ can
be obtained by averaging along a long KMC trajectory:
\begin{equation}
\mathcal{\left\langle P\right\rangle }=\dfrac{\sum_{m=1}^{K}\tau_{m}\mathcal{P}_{m}}{\sum_{m=1}^{K}\tau_{m}},\label{eq:average}
\end{equation}
where $\mathcal{P}_{m}$ is the value of the property in the microstate
visited at the Monte Carlo step $m$, and $K\gg1$ is the number of
Monte Carlo steps on the trajectory. Assuming ergodicity, $\mathcal{\left\langle P\right\rangle }$
approximates the ensemble average of the property.

\section{Equilibrium properties of the solute-free system}

Before analyzing the GB-solute interactions, we applied the KMC algorithm
to study the equilibrium properties of a solute-free system. The goal
was threefold: (1) validate our methodology, (2) calculate the equilibrium
properties of solute-free GBs for the subsequent comparison with those
in the alloy system, and (3) generate reference data required for
thermodynamic integration in the alloy systems. In the absence of
solute atoms, the KMC simulations only implemented the flip transitions.

\subsection{The bulk transition temperature}

\begin{figure}[!htb]
\centering{}\centering \includegraphics[width=0.38\textwidth]{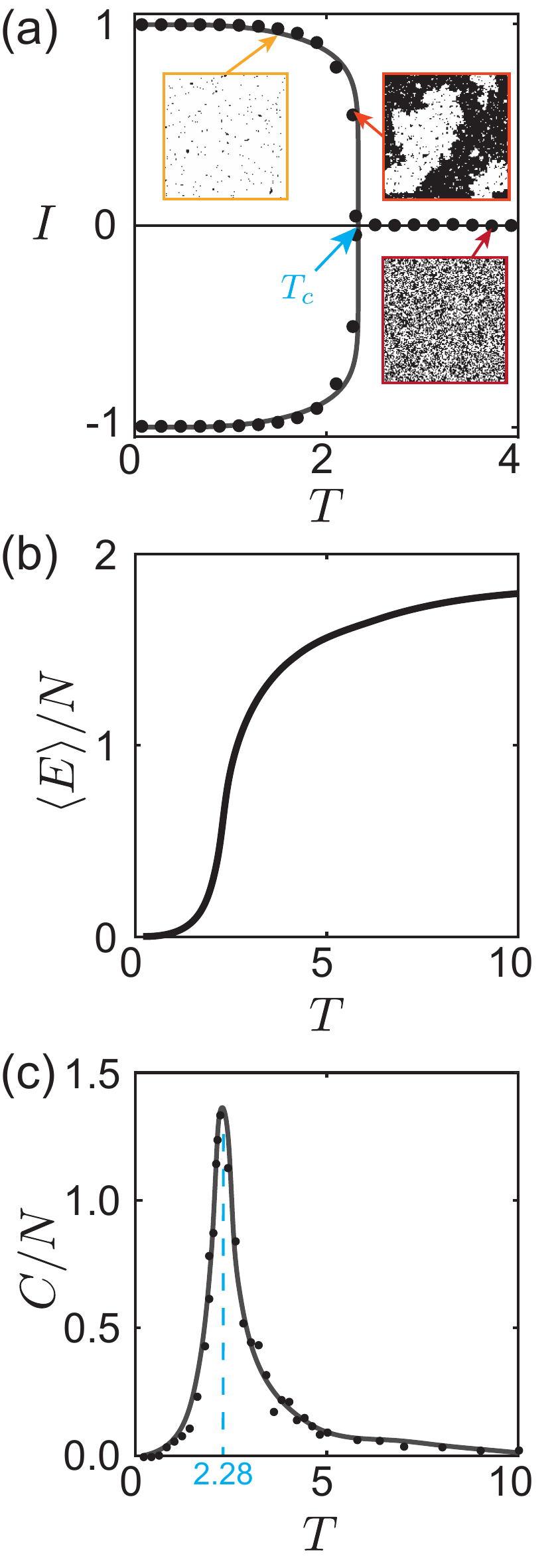}
\caption{Properties of the solute-free system. (a) Average orientation (color)
$I$ as a function of temperature. The points were computed in this
work and the lines represent the analytical solution (\ref{eq:Yang-1952}).
The insets show typical structures of single-crystalline states at
low temperatures, near the critical point $T_{c}$, and the disordered
state above $T_{c}$. (b) Energy per cell as a function of temperature.
The curve has been smoothed for numerical differentiation. (c) Heat
capacity per cell computed by Eqs.(\ref{eq:C1}) (curve) and (\ref{eq:C2})
(points). The peak at $T=2.28$ marks the approximate location of
the critical temperature $T_{c}$. }
\label{fig:total_energy-1}
\end{figure}

The bulk transition temperature was computed by a set of KMC simulations
on a $128\times128$ system at temperatures $T$ ranging from 0.1
to 10 with 0.1 increments. At each temperature, the simulation started
with a single crystalline state and was run for about $10^{7}$ Monte
Carlo steps after the system reached equilibrium. The average color
of the system is plotted as a function of temperature in Fig.~\ref{fig:total_energy-1}(a).
The heat capacity per atom, $C$, was calculated using two methods.
First, the total energy $\left\langle E\right\rangle $ was computed
from Eq.(\ref{eq:average}) as a function of $T$, and the heat capacity
was obtained by numerical differentiation:
\begin{equation}
C=\dfrac{d\left\langle E\right\rangle }{dT}.\label{eq:C1}
\end{equation}
The second method was based on the fluctuation formula \citep{Landau-Lifshitz-Stat-phys,Mishin:2015ab}
\begin{equation}
C=\dfrac{\left\langle E^{2}\right\rangle -\left\langle E\right\rangle ^{2}}{T^{2}}.\label{eq:C2}
\end{equation}

As shown in Fig.~\ref{fig:total_energy-1}(b,c), both methods give
similar results, confirming the robustness of our algorithm. The heat
capacity peak corresponds to the bulk phase transition, which allows
us to estimate the transition temperature $T_{c}$ at $2.28$. For
comparison, in the 2D Ising model on a square lattice, the exact transition
temperature derived by Onsager~\citep{Onsager:1944aa} is 
\begin{equation}
T_{c}=\frac{2}{\ln\left(1+\sqrt{2}\right)}\approx2.269.\label{Eqn:BTT}
\end{equation}
The slight discrepancy could be reduced using a larger system and
more temperatures, which we did not pursue.

At high temperatures approaching $T_{c}$ from below, the initial
grain ($I=\pm1$) develops a high concentration of randomly distributed
cells of opposite color. At the bulk transition temperature, the system
becomes a random mixture of black and white cells in equal proportion
($I\rightarrow0$). As shown in Fig.~\ref{fig:total_energy-1}(a),
the simulation results closely follow the analytical solution \citep{Yang:1952aa}
\begin{equation}
I=\left[1-\mathrm{csch}^{2}\left(\frac{2}{T}\right)\right]^{1/8}.\label{eq:Yang-1952}
\end{equation}

In magnetic systems, $T_{c}$ represents the paramagnetic transition
temperature. Given our interpretation of the cells as representing
local lattice orientations, we can interpret $T_{c}$ as the ``melting''
temperature of the single crystal. Note that a polycrystalline structure
with equal numbers of black and white cells can also give $I=0$ at
temperatures below $T_{c}$. However, this structure is thermodynamically
unstable: solute-free GBs in this model possess a positive excess
free energy, which creates a driving force for grain coarsening. The
coarsening continues until the system transforms into a single crystal
with $I=\pm1$. As shown later, solute atoms can stabilize GBs and
create thermodynamically equilibrium polycrystalline structures below
$T_{c}$.

\subsection{GB free energy from thermodynamic integration\label{subsec:gamma_pure}}

Thermodynamic properties of solute-free GBs were studied using a system
of two GBs parallel to the $x$-axis. The system size in the direction
normal to the GBs was fixed at $Y=256$. Three system sizes parallel
to the GBs were tested, with the $X$ values of $128$, $256$, and
$512$. To ensure that the system has reached thermodynamic equilibrium
and collect sufficient statistics of capillary waves, the simulations
comprised about $10^{9}$ Monte Carlo steps.

The interfacial free energy $\gamma$ was calculated by the thermodynamic
integration method reviewed by Binder et al.~\citep{Binder2011}.
The method is based on the interfacial form of the Gibbs-Helmholtz
equation,
\begin{equation}
\frac{\partial(\tilde{F}/T)}{\partial T}=-\frac{\tilde{E}}{T^{2}},\label{eq:GH-1}
\end{equation}
where $\tilde{E}$ and $\tilde{F}$ are the excess GB energy and free
energy, respectively. For a 2D system with periodic boundaries,
\begin{equation}
\tilde{F}=\gamma L-TS,\label{eq:F}
\end{equation}
where $L$ is the GB length. The latter generally differs from its
geometric length $X$ due to the capillary waves. In Eq.(\ref{eq:F}),
$S$ is the configurational entropy of the GB, which takes into account
that the GB can be located at any position along the $y$-axis without
changing the system energy. For a single GB, $S=\ln Y$. Since we
have two GBs, their configurational entropy is $S=\ln(Y^{2}/2)$,
where we assumed that the GBs can choose their positions independently
but swapping their positions does not create a new state (hence the
division by $2$). Thus, the configuration entropy per one GB is 
\begin{equation}
S=\dfrac{1}{2}\ln\left(\dfrac{Y^{2}}{2}\right).\label{eq:S}
\end{equation}

Eq.(\ref{eq:GH-1}) is integrated with respect to temperature from
a chosen reference temperature $T_{0}$ to the current temperature
$T$:
\begin{equation}
\gamma=\gamma_{0}\dfrac{XT}{LT_{0}}-\dfrac{T}{L}\int_{T_{0}}^{T}\frac{\tilde{E}}{T^{\prime2}}\,dT^{\prime}+\dfrac{T}{L}\left[\ln\left(\dfrac{Y}{\sqrt{2}}\right)-S_{0}\right],\label{eq:gamma}
\end{equation}
where $\gamma_{0}$ is the GB free energy at the temperature $T_{0}$.
The reference temperature is chosen low enough to disregard the entropy
effects and treat the GB as perfectly straight ($L=X$). At this temperature,
we can take $\gamma_{0}X=\tilde{E}_{0}$ and neglect the configurational
entropy $S_{0}$. 

\begin{figure}[!htb]
\centering{}\centering \includegraphics[width=1\linewidth]{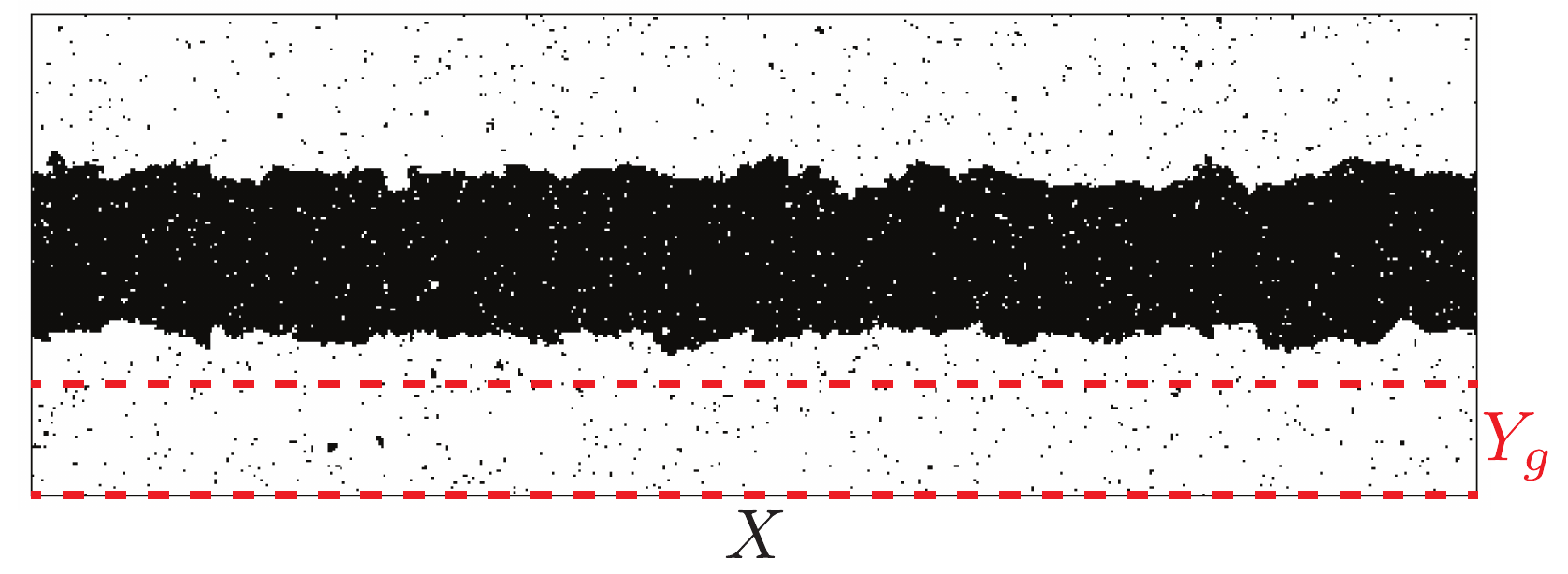}
\caption{The excess GB energy calculations by Eq.(\ref{eq:Excess}). The dashed
rectangle outlines a region with dimensions $X\times Y_{g}$ selected
to represent the grain energy $E_{g}$. }
\label{fig:IsingModelMethods}
\end{figure}

The excess energy $\tilde{E}$ was calculated as follows:
\begin{equation}
\tilde{E}=\dfrac{1}{2}\left\langle E-\dfrac{Y}{Y_{g}}E_{g}\right\rangle .\label{eq:Excess}
\end{equation}
The angular brackets denote the time averaging as in Eq.(\ref{eq:average}),
and the factor of $1/2$ is due to the presence of two GBs in the
system. In Eq.(\ref{eq:Excess}), $E$ is the total energy of the
system and $E_{g}$ is the energy of a single-crystalline region unaffected
by the GBs. This region was chosen inside one of the grains with dimensions
$X\times Y_{g}$ (Fig.~\ref{fig:IsingModelMethods}). 

We chose $T_{0}=0.3$, computed $\tilde{E}(T)$ on a temperature grid,
and performed a quadratic spline interpolation for numerical integration.
The GB free energy obtained is plotted as a function of temperature
in Fig.~\ref{fig:IsingModel_Interface}(a). The curve agrees well
with previous Metropolis Monte Carlo calculations within the Ising
model \citep{Binder2011}. In particular, the GB free energy decreases
with temperature and converges to zero at the bulk critical temperature
$T_{c}$. At this temperature, the difference between the grain orientation
vanishes, and the GB ceases to exist. 

\subsection{GB stiffness calculation\label{subsec:stiffness}}

In the previous section, the GB free energy was calculated for the
GB orientation parallel to the $x$-axis. However, GB properties in
this model are anisotropic. This anisotropy can be approximately represented
by the GB stiffness $\alpha$ defined by
\begin{equation}
\alpha=\gamma_{||}+(\partial^{2}\gamma/\partial\theta^{2})_{||},\label{eq:3.2-1}
\end{equation}
where $\theta$ is the inclination angle of the GB relative to the
$x$-axis. The symbol $||$ indicates that both $\gamma$ and its
second angular derivative refer to the $\theta\rightarrow0$ limit. 

We have computed the GB stiffness as a function of temperature by
the capillary fluctuation method \citep{Hoyt01,Morris02,Pun:2020aa,Mishin2014}.
In this method, the fluctuated GB shape is decomposed into a Fourier
series with amplitudes $A_{n}$. The mean-square Fourier amplitude
is related to the GB stiffness by the equation (rewritten in our dimensionless
variables)
\begin{equation}
\langle|A_{n}|^{2}\rangle=\frac{T}{\alpha Xk_{n}^{2}},\label{eq:CFM}
\end{equation}
where $k_{n}=(2\pi/X)n$ ($n=\pm1,\pm2,..$) are wave numbers. Thus,
$\alpha$ can be extracted from the slope of the plot $\langle|A_{n}|^{2}\rangle^{-1}$
versus $k_{n}^{2}$ in the long-wave (small $k$) limit.

\begin{figure}[!htb]
\centering \includegraphics[width=0.8\linewidth]{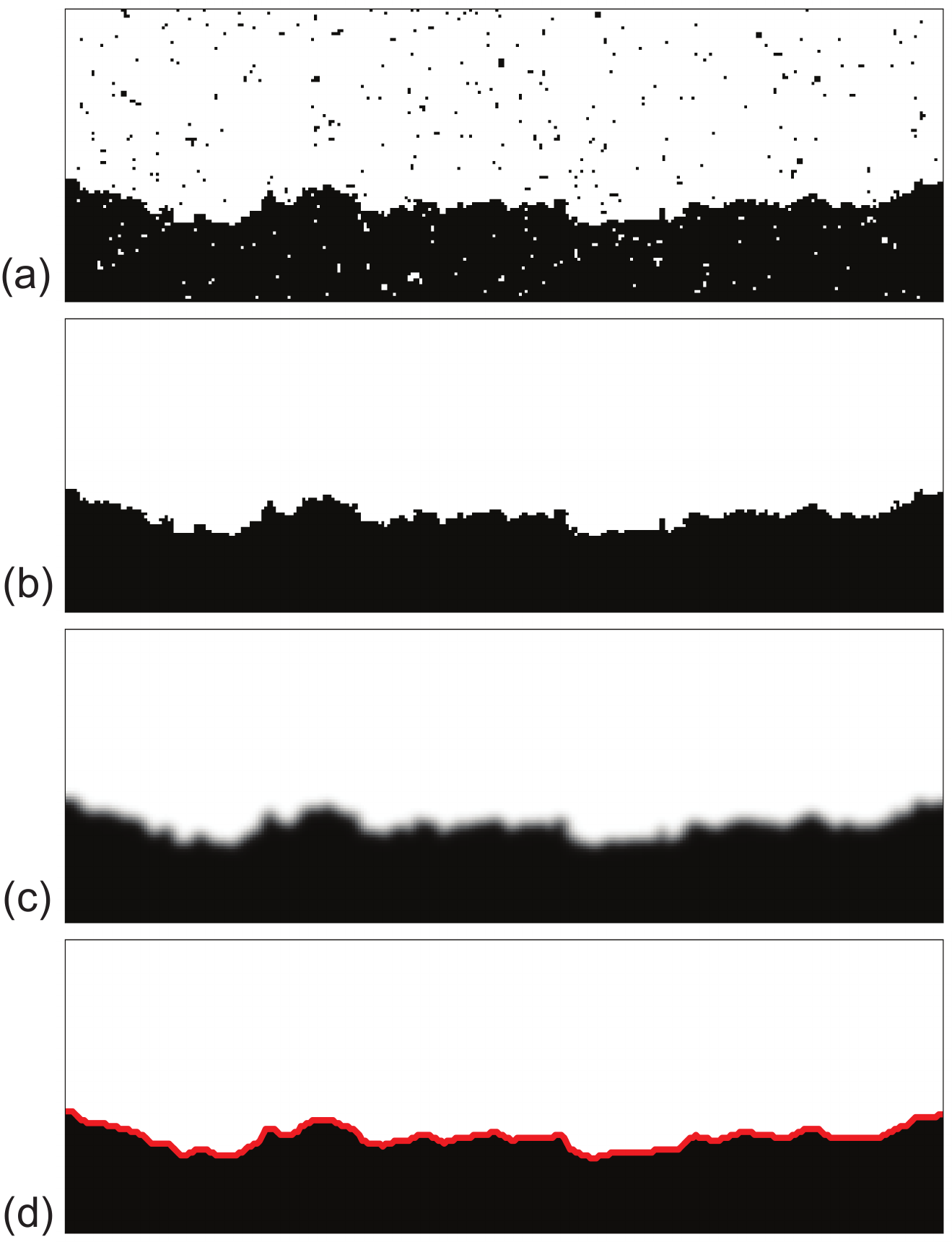}
\caption{Example of GB shape calculation in the capillary fluctuation method.
(a) Initial structure containing a fuzzy GB and background noise in
the grains in the form of opposite-color cells. (b) The structure
after removing the background noise in the grain. (c) GB shape smoothed
by a Gaussian filter. (d) Final smoothed GB shape shown in red.}
\label{fig:StiffnessMethod}
\end{figure}

The critical step in the capillary fluctuation method is to extract
the GB shape from the simulation snapshots. The challenge is caused
by the fuzziness of the GBs and the background noise in the grains.
In our case, the grains contain numerous cells of the opposite color,
either isolated or aggregated into small clusters reflecting the short-range
order (Fig.~\ref{fig:StiffnessMethod}(a)). This background noise
was especially significant at high temperatures. An algorithm was
developed to identify the ``wrong'' color cells/clusters lying outside
the GBs and reverse their color to that of the grain (Fig.~\ref{fig:StiffnessMethod}(b)).
This grain-cleaning procedure was applied to all snapshots saved during
the simulations. Next, the GBs were smoothed by a Gaussian filter
whose width was adjusted to preserve the GB shape on a length scale
moderately exceeding the cell size (Fig.~\ref{fig:StiffnessMethod}(c)).
Finally, the smoothed GB shape was traced ensuring a one-to-one correspondence
(Fig.~\ref{fig:StiffnessMethod}(d)). 

The discrete Fourier transformation was applied to the smoothed GBs,
and the Fourier amplitudes obtained were averaged over a set of KMC
snapshots. The $\alpha$ values extracted from the $\langle|A_{n}|^{2}\rangle^{-1}$
versus $k_{n}^{2}$ plots are shown in Fig.~\ref{fig:total_energy-1}
as a function of temperature. The results agree well with previous
Metropolis Monte Carlo simulation within the Ising model \citep{Binder2011}.
Like the GB free energy $\gamma$, the stiffness converges to zero
at the bulk critical temperature $T_{c}$. At lower temperatures,
the stiffness deviates from $\gamma$, indicating that the torque
term in Eq.(\ref{eq:3.2-1}) becomes significant. 

Note that $\alpha$ diverges to infinity at $T\rightarrow0$. In this
limit, the GB is nearly straight with a small concentration of kinks
\citep{Gelfand:1990vz,Lapujoulade:1994uw,Saito:1996vo}. This GB structure
creates a cusp in the angular dependence of $\gamma$ with a diverging
second derivative. Although these calculations validate our methodology,
the applicability of the capillary fluctuation method at low temperatures
is questionable because the capillary wave amplitudes become indistinguishable
from the intrinsic GB width. The difference between the intrinsic
and capillary GB widths is discussed next.

\subsection{One GB, two widths\label{subsec:widths}}

\begin{figure}[!htb]
\centering \includegraphics[width=0.6\linewidth]{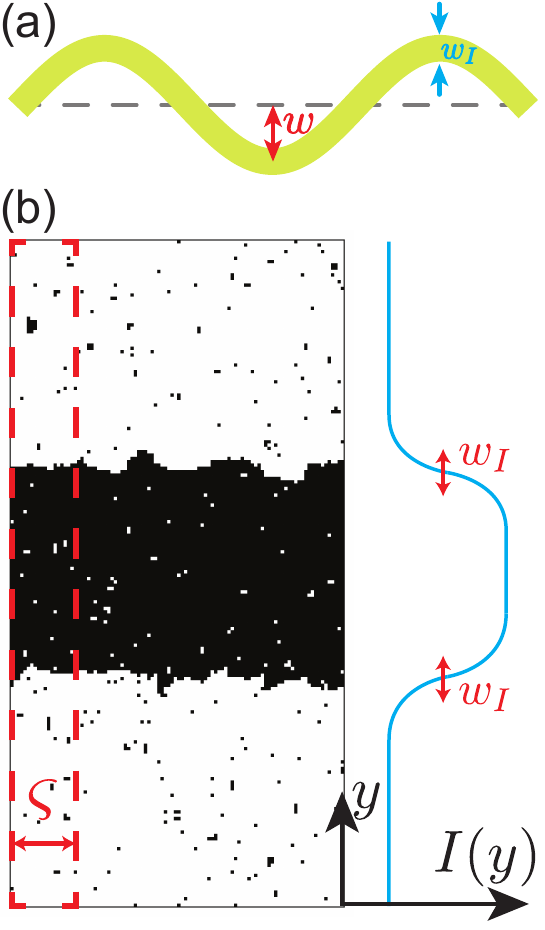}
\caption{(a) Definition of the intrinsic GB width $w_{I}$ and the mean-squared
capillary width $\left\langle w^{2}\right\rangle $. (b) Example of
the $w_{I}$ calculation. The dashed box indicates a stripe normal
to the GB. The blue curves show the hyperbolic functions fitted to
the color function $I(y)$ in the GB regions.}
\label{fig:IntrinsicInterfaceWidthMethod-1}
\end{figure}

A GB can be characterized by two widths, which differ both conceptually
and numerically (Fig.~\ref{fig:IntrinsicInterfaceWidthMethod-1}(a)).
The intrinsic GB width $w_{I}$ is the width of the atomic-scale transition
layer between the grains. The capillary GB width characterizes the
amplitude of the capillary waves on a larger length scale. It is customary
to quantify the capillary width by the mean-squared deviation $\left\langle w^{2}\right\rangle $
from the average GB plane (line in 2D):
\begin{equation}
\left\langle w^{2}\right\rangle \equiv\dfrac{1}{X}\left\langle \sum_{n=1}^{X}(y_{n}-\overline{y})^{2}\right\rangle .\label{eq:2.04}
\end{equation}
Here, $y_{n}$ are $y$-coordinates of the GB cells and 
\[
\overline{y}=\dfrac{1}{X}\sum_{n=1}^{X}y_{n}
\]
is the instantaneous GB position.

Both $w_{I}$ and $\left\langle w^{2}\right\rangle $ increase with
temperature and diverge to infinity at the bulk critical point. At
a given temperature, $w_{I}$ and $\left\langle w^{2}\right\rangle $
follow different system size dependencies. $w_{I}$ is an intensive
GB property and does not depend on the system size. By contrast, $\left\langle w^{2}\right\rangle $
increases with the GB length $X$. Indeed, according to the capillary
wave theory \citep{Gelfand:1990vz,Lapujoulade:1994uw,Saito:1996vo,Mishin:2023ab},
\begin{equation}
\left\langle w^{2}\right\rangle =\dfrac{TX}{12\alpha}.\label{eq:3.12-1}
\end{equation}
This equation predicts a linear scaling with $X$.

In the simulations, $\left\langle w^{2}\right\rangle $ could be readily
calculated from the GB shapes discussed in the previous section. To
calculate $w_{I}$, we developed a procedure illustrated in Fig.~\ref{fig:IntrinsicInterfaceWidthMethod-1}(b).
The system was divided into thin slices parallel to the $y$-axis.
The $x$-dimension of each slice was $\varsigma\ll X$, so the GB
portion within each slice could be treated as linear (no capillary
waves). We found that $\varsigma=4$ was a reasonable choice. Each
slice was further divided into thin layers parallel to the $x$-axis,
and the color $I$ of each layer was computed by averaging the $\sigma$-numbers
of the cells using Eq.(\ref{eq:0}). This calculation produced a function
$I(y)$ for each slice. This function was relatively close to $1$
or $-1$ within the grains and transitioned between these values across
the GBs. In the transition region, this function was fitted by the
hyperbolic tangent with three fitting parameters $I_{0}$, $y_{0}$
(the local GB position), and $w_{I}$, 
\begin{equation}
I(y)=I_{0}\tanh\left(\dfrac{y-y_{0}}{w_{I}}\right),\label{eq:tanh}
\end{equation}
by minimizing the mean-squared deviation. The final result for $w_{I}$
was obtained by averaging over the two GBs, all slices, and all snapshots
saved during the simulations.

\begin{figure}[!htb]
\centering{}\centering \includegraphics[width=0.8\textwidth]{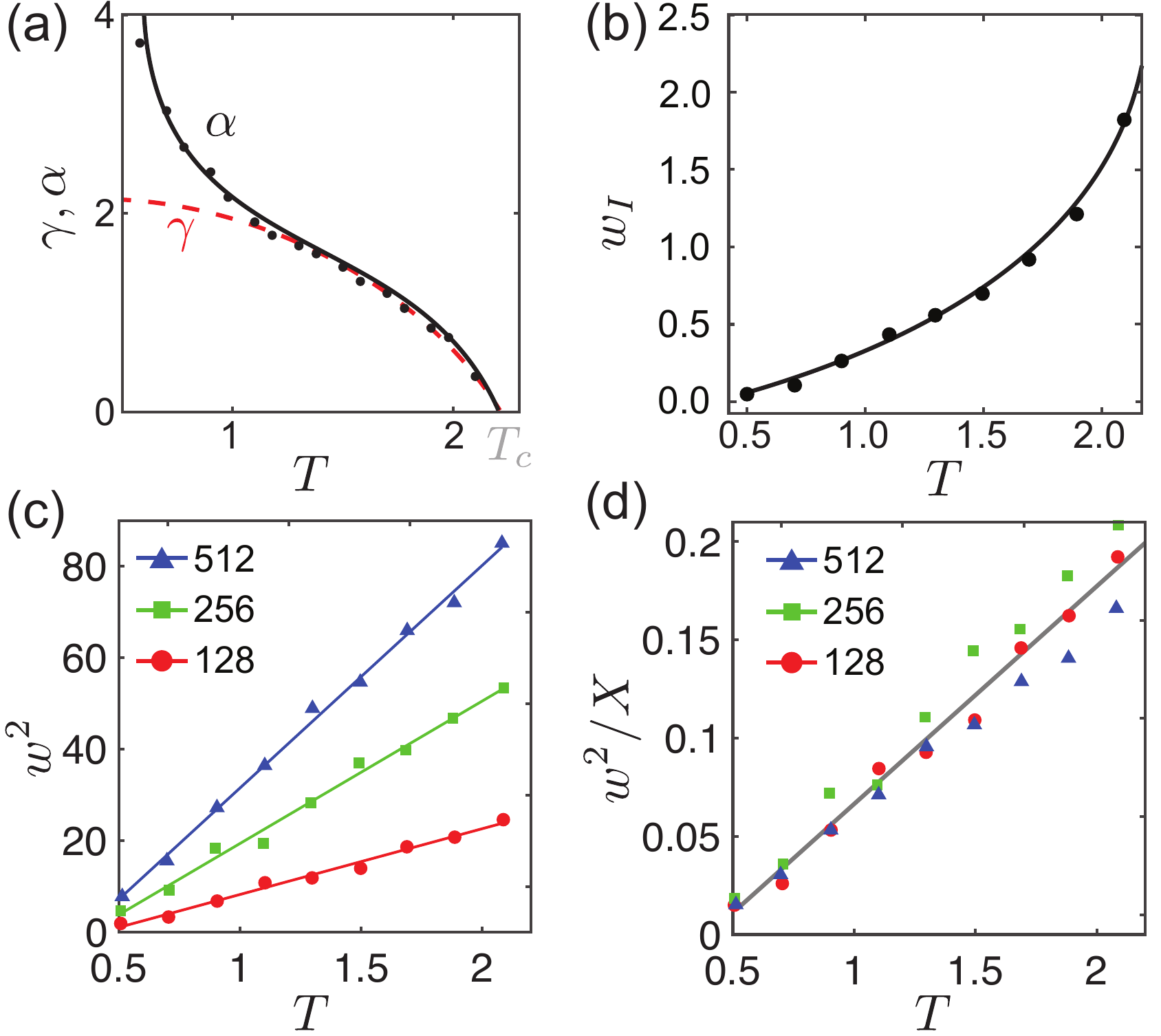}
\caption{Solute-free GB properties as a function of temperature. (a) GB free
energy $\gamma$ and stiffness $\alpha$. The points represent individual
calculations of $\alpha$ by the capillary wave method, while the
curve serves as a guide to the eye. (b) Intrinsic interface width
$w_{I}$. (c) Mean-squared GB width $w^{2}$ for three system sizes
$X$ indicated in the key. (d) Normalized mean-squared interface width
$w^{2}/X$ for three system sizes $X$ indicated in the key. The highest
temperature shown on the plots is close to the bulk critical point
$T_{c}$. The results were obtained by KMC simulations within the
present model.}
\label{fig:IsingModel_Interface}
\end{figure}

Fig.~\ref{fig:IsingModel_Interface}(b,c) shows the obtained temperature
dependencies of $w_{I}$ and $\left\langle w^{2}\right\rangle $.
As expected, both widths increase with temperature, reflecting the
GB broadening as the average grain orientations (colors) converge
at high temperatures. Note that the temperature dependencies of $\left\langle w^{2}\right\rangle $
are fairly linear for all three system sizes tested, as predicted
by Eq.(\ref{eq:3.12-1}). This equation also predicts that the plots
for different system sizes must collapse into a single straight line
when plotted in the $\left\langle w^{2}\right\rangle /X$ versus $T$
format. Fig.~\ref{fig:IsingModel_Interface}(d) shows that this prediction
is borne out by the simulations. The remaining scatter is due to the
statistical fluctuations increasing with temperature. Comparison of
Fig.~\ref{fig:IsingModel_Interface}(b) and (c) indicates that the
capillary GB width $\sqrt{\left\langle w^{2}\right\rangle }$ is significantly
larger than the intrinsic width $w_{I}$ at $T>0.5$, justifying the
validity of the capillary fluctuation method at these temperatures.

\section{Bulk thermodynamics of alloys}

Having tested the thermodynamic properties of the solute-free system,
we are now ready to investigate the alloy systems. Two computational
methods will be applied. To study the equilibrium thermodynamics,
we use the Metropolis Monte Carlo method described in section \ref{subsec:Metropolis}.
We consider a $256\times128$ system connected to a thermostat at
a temperature $T$ and to an imaginary reservoir of solute atoms at
a chemical potential $\mu$. Upon equilibration, the system reaches
a solute concentration $c$ and an average orientation (color) $I$.
Different equilibrium states can be explored by varying the control
parameters $T$ and $\mu$. The results can be represented by phase
diagrams in the coordinates $T$, $c$ and $I$ (note, however, that
only two of them are independent variables). Having the phase diagrams,
we then apply KMC simulations for a more detailed investigation of
the equilibrium structures at different locations on the phase diagram.
The KMC barriers $\varepsilon_{0}=1$, $\varepsilon_{00}^{s}=1$,
and $q=0$ are chosen, but this choice does not play a role because
we are only interested in equilibrium structures. The choice of the
barriers will become important when we examine the role of solute
diffusion in Section \ref{sec:Solute-segregation}.

Different choices are possible for the solute-GB interaction energy
$J_{sg}$ and the solute-solute interaction energy $J_{ss}$. If the
latter is positive, the alloy can separate into two phases with different
solute concentrations below a critical point that is different from
the orientational critical temperature $T_{c}$. To simplify the parametric
analysis, we will focus on the case of $J_{ss}=0$. Accordingly, only
variations in $J_{sg}$ are considered.

We will assume that $J_{sg}>0$ to ensure attraction between solute
atoms and cells with mixed (black and white) environments. This attraction
causes solute segregation to GBs, but it also produces the reverse
effect, in which the solute atoms inside the grains tend to surround
themselves with cells of opposite color relative to the color of the
grain. In other words, the solute atoms amplify short-range disorder
by promoting the formation of clusters of misoriented cells centered
at a solute atom. The enhanced disorder can be expected to reduce
the bulk critical point relative to the solute-free case. The misoriented
cells surrounding a solute atom can be interpreted as lattice distortions
existing around solute atoms in real crystals.

\begin{figure}[!htb]
\centering{}\centering \includegraphics[width=0.72\linewidth]{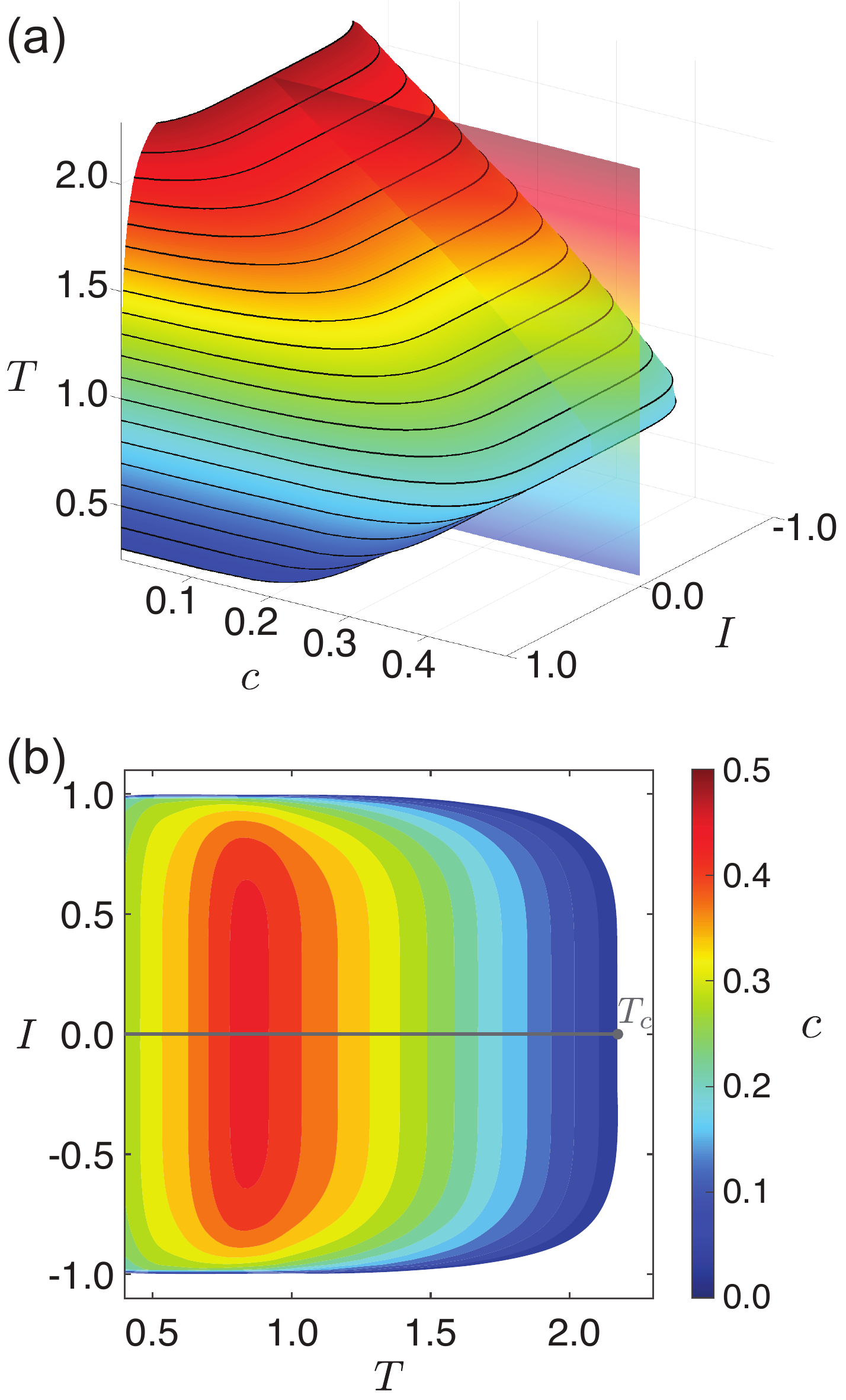}
\caption{(a) Phase diagram in the composition-temperature-orientation coordinates.
The segregation energy is fixed at $J_{sg}=2$. (b) Slices of the
phase diagram at different chemical compositions indicated in the
color bar. Note that the iso-concentration contours form closed loops
as the solute concentration increases. These loops indicate the existence
of two critical points. }
\label{fig:PhaseDiagram}
\end{figure}

The surface plot in Fig.~\ref{fig:PhaseDiagram}(a) is the $T-c-I$
phase diagram computed with $J_{sg}=2$ (recall that $J_{sg}$ is
normalized by $J_{gg}$). The surface is colored according to temperature,
and the lines on the surface are $c-I$ isotherms. In the $c\rightarrow0$
limit, the ends of each isotherm converge to crystalline states with
$I$ values close to $1$ or $-1$. As the solute concentration increases
at a fixed temperature, the positive and negative $I$ values decrease
in magnitude, indicating that the grains develop significant orientational
disorder (as predicted above). At a critical solute concentration,
both $I$ values converge to zero, and the system becomes orientationally
disordered. At this chemical composition, the chosen temperature becomes
the critical temperature above which a crystalline state cannot exist.

The $I-T$ plot in Fig.~\ref{fig:PhaseDiagram}(b) represents a set
of iso-concentration cross-sections of the phase diagram. The $c=0$
cross-section reproduces the solute-free diagram shown in Fig.~\ref{fig:total_energy-1}(a).
When the solute concentration becomes high enough, the iso-concentration
cross-sections form closed loops. Such loops indicate that single
crystalline states of the alloy only exist in a certain temperature
interval. In other words, the alloy system has \emph{two} critical
points. The upper critical temperature smoothly connects to the solute-free
$T_{c}$ in the $c\rightarrow0$ limit and represents a melting transition.
The lower critical point exists only in alloys and represents a phenomenon
in which a single crystal loses thermodynamic stability upon cooling
below this temperature. 

\begin{figure}
\begin{centering}
\includegraphics[width=1\textwidth]{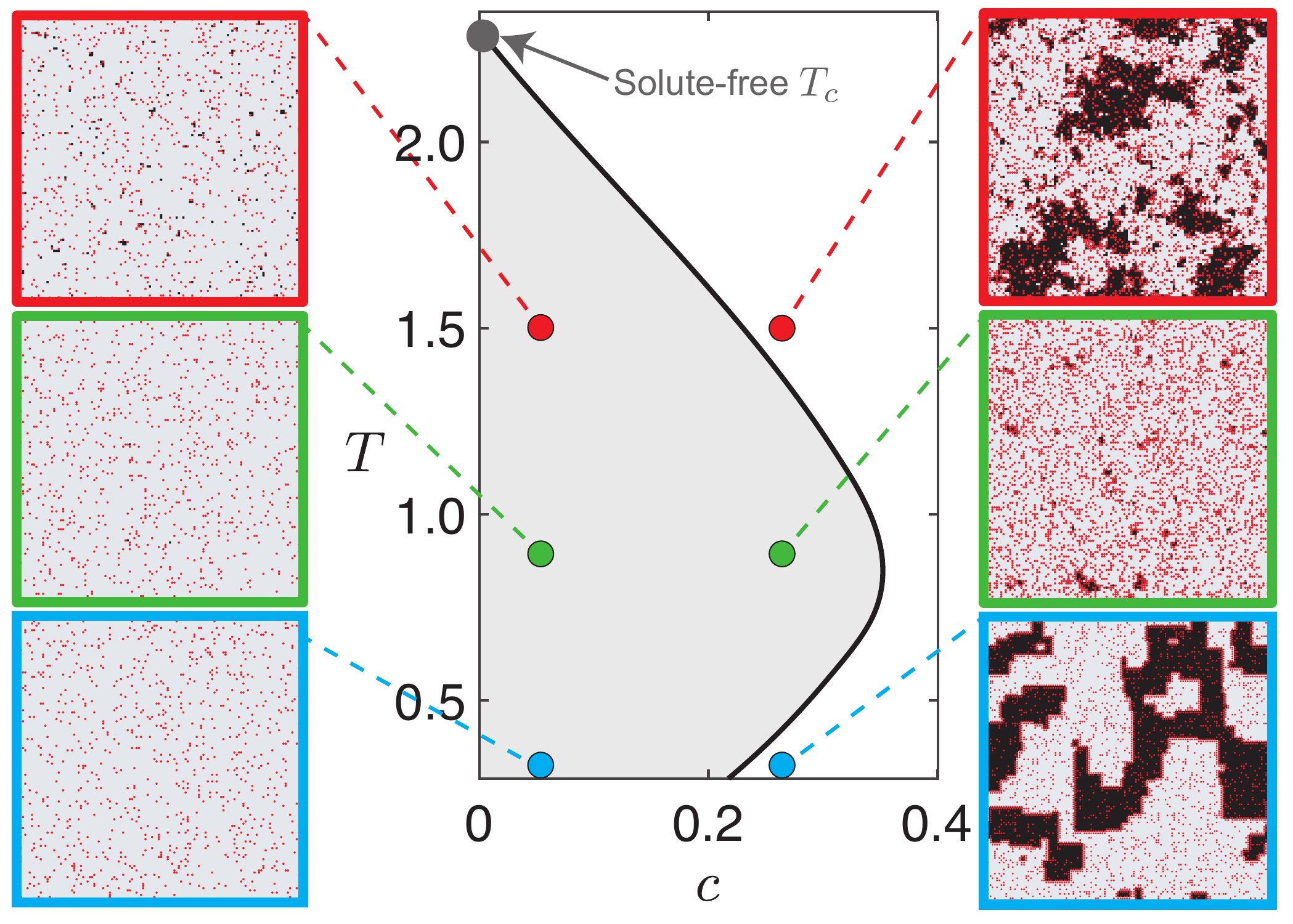}
\par\end{centering}
\caption{Temperature-composition phase diagram of alloys. The insets show representative
structures at selected points on the diagram. The segregation energy
is $J_{sg}=2$. The red dots represent solute atoms. \label{fig:CriticalTemperature}}

\end{figure}

The existence of the low critical temperature is also evident from
the nose-like shape of the surface in Fig.~\ref{fig:PhaseDiagram}(a),
in which the states under the nose (shown in blue) represent the low
critical points. To further clarify the meaning of such points, we
plot the critical temperature as a function of alloy composition in
Fig.~\ref{fig:CriticalTemperature}. This plot is the intersection
of the phase diagram in Fig.~\ref{fig:PhaseDiagram}(a) with the
$I=0$ plane. It shows that the critical temperature is a two-valued
function of the chemical composition, with the upper and lower branches
representing two critical lines. The upper branch reflects the effect
of critical point suppression by alloying mentioned above. The lower
branch is the boundary of single-crystal stability at low temperatures.
Single-crystalline states of either orientation only exist in the
gray area on the left of the critical curve. 

The insets in Fig.~\ref{fig:CriticalTemperature} display representative
structures of the system in several parts of the phase diagram. These
structures were obtained by closing the system (no solute exchange
with the reservoir) and equilibrating it by KMC simulations. The images
on the left show the structures of a white grain at the temperatures
of 0.3, 0.9, and 1.5 with the fixed chemical composition of $c=0.05$.
Note the local disorder in the form of single and clustered black
cells, which increases with temperature and becomes especially visible
at $T=1.5$. 

The images on the right correspond to the same three temperatures
but with a solute concentration of $c=0.25$. At $T=0.9$, the system
is on the left of the critical line, so the single-crystalline state
is stable. Compared to the $c=0.05$ structure at the same temperature,
more extensive clustering is observed due to the solute effect. At
$T=1.5$, single-crystalline states are unstable and the system is
orientationally disordered with $I=0$. Because this liquid-like state
is relatively close to the critical line, it exhibits large orientational
fluctuations on the scale comparable to the system size. Significant
clustering into black and white regions is observed, in contrast to
the homogeneous supercritical structure in the solute-free system
(cf.~Fig.~\ref{fig:total_energy-1}(a)). Such regions are dynamic:
they constantly form, dissolve, and move around while preserving the
zero average orientation of the system ($I=0$). Finally, the structure
equilibrated at $T=0.3$ is composed of back and white grains separated
by sharp GBs with significant solute segregation. Note that the GBs
are faceted, confirming that the structure is crystalline in contrast
to the liquid-like state at $T=1.5$. We interpret this structure
as a thermodynamically stable polycrystalline state discussed in section
\ref{sec:Intro}. Since our model only considers two crystallographic
orientations, the large GB area is achieved by forming enclosed grains,
with black grains nested inside the white and vice versa.

To further demonstrate the different equilibrium states mentioned
above, we created a circular black grain inside a white parent grain,
as shown in Fig.~\ref{fig:grain-evolution}. In the initial state,
a uniform random distribution of solute atoms with $c=0.25$ was created.
The system was then equilibrated by KMC simulations at the temperatures
of 0.3, 0.9, and 1.5. The initial, intermediate, and equilibrated
structures obtained are shown in three columns. At $T=1.5$, the black
grain disintegrates, forming a disordered, liquid-like structure similar
to that in Fig.~\ref{fig:CriticalTemperature}. This behavior confirms
that a crystalline state is unstable at this temperature. At $T=0.9$,
the crystalline state is expected to be stable according to the phase
diagram. And indeed, both grains remain crystalline but the black
grain shrinks and soon disappears. This scenario indicates that the
GB free energy is positive, creating a capillary driving force causing
the grain shrinkage. Finally, at $T=0.3$, the black grain does not
shrink. Instead, it breaks into fragments and eventually forms a fine-grained
structure with faceted GBs similar to the structure observed in Fig.~\ref{fig:CriticalTemperature}.
The fragmentation process increases the total GB area until the system
arrives at thermodynamic equilibrium with a finite average grain size.
In this structure, the GB free energy is expected to be zero.

\begin{figure}
\centering{}\includegraphics[width=0.57\textwidth]{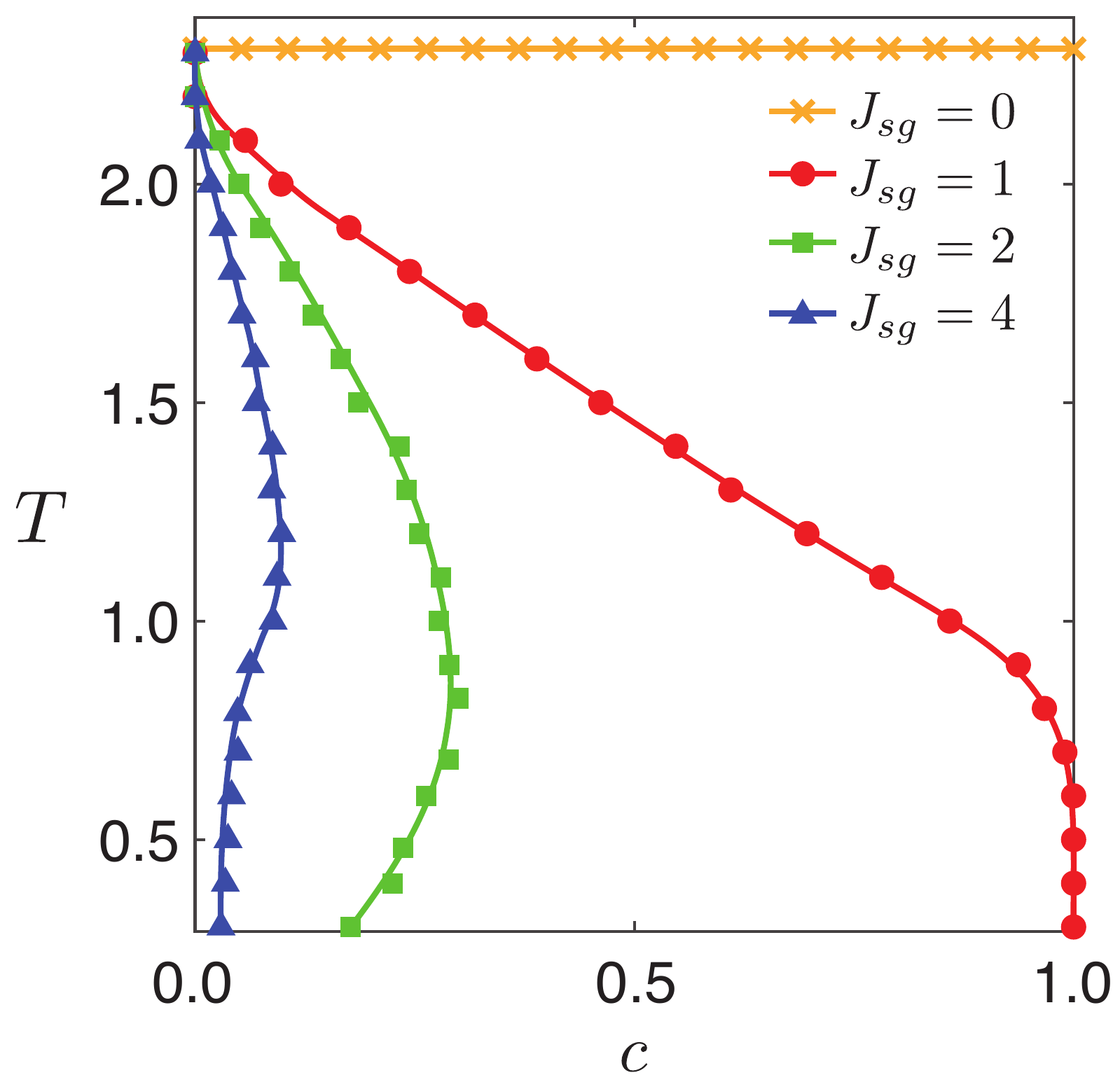}\caption{Temperature-composition phase diagrams for different choices of the
segregation energy $J_{sg}$ indicated in the key. The points represent
individual simulations by the Metropolis Monte Carlo algorithm. \label{fig:J_sg-effect}}
\end{figure}

The results shown above were obtained with the segregation energy
of $J_{sg}=2$. The effect of this energy on the phase diagram is
illustrated in Fig.~\ref{fig:J_sg-effect}. At larger $J_{sg}$ values,
the nose shape of the critical line leading to the formation of stable
polycrystals continues to exist. Furthermore, the polycrystalline
state can be stabilized at smaller solute concentrations. As $J_{sg}$
decreases, the critical line shifts towards higher concentrations
and straightens until the GB stabilization effect disappears. In the
$J_{sg}\rightarrow0$ limit, polycrystalline states are less stable
than a single crystal for any concentration, including the pure-solute
system ($c=1$).

\section{Grain boundary free energy in alloys\label{sec:TDI}}

To validate the link between the $\gamma=0$ condition and the polycrystalline
stability established in the previous section, we performed direct
calculations of the GB free energy. The starting point was the 2D
form of the Gibbs adsorption equation
\begin{equation}
d(\gamma L)=-\tilde{N}d\mu,\enskip\enskip T=\mathrm{const},\label{eq:adsorption}
\end{equation}
where $\tilde{N}$ is the total amount of GB segregation. The latter
is defined by
\begin{equation}
\tilde{N}=(c-c_{g})N/2,\label{eq:segreg}
\end{equation}
where $c_{g}$ is the solute concentration in grain regions unperturbed
by the GBs, and $c$ is the average solute concentration in the system.
As before, the factor of two accounts for the presence of two GBs. 

Integrating Eq.(\ref{eq:adsorption}), we obtain 
\begin{equation}
\gamma(T,\mu)=\dfrac{1}{L(T,\mu)}\left[\gamma(T,\mu_{*})L(T,\mu_{*})-\intop_{\mu_{*}}^{\mu}\tilde{N}(T,\mu^{\prime})d\mu^{\prime}\right].\label{eq:Adsorption_integrated}
\end{equation}
We display all variables involved in the integration for the sake
of clarity. The asterisk subscript refers to the reference state $(T,\mu_{*})$.
Recall that $L$ is the ensemble-averaged GB length, which can differ
from the system size $X$ due to capillary fluctuations. In practice,
the reference chemical potential $\mu_{*}$ is chosen to achieve the
smallest possible solute concentration at which $\tilde{N}$ can still
be reliably extracted from the simulations. Then $\gamma(T,\mu_{*})$
and $L(T,\mu_{*})$ are approximated by their solute-free values at
the temperature $T$. The solute-free $\gamma(T)$ values were calculated
in section \ref{subsec:gamma_pure} by thermodynamic integration with
respect to temperature. Technical details of computing the ensemble-averaged
values of $\tilde{N}$ and $L$ and performing the numerical integration
are presented in the online Supplementary Information file accompanying
this article.

Metropolis Monte Carlo simulations were performed on a $128\times256$
system with two GBs parallel to the $x$-axis. Temperature was varied
between 0.3 and 2.0, and the GB free energy $\gamma(T,\mu)$ was calculated
from Eq.(\ref{eq:Adsorption_integrated}) at each temperature. The
results were converted to the function $\gamma(T,c)$ using the relationship
between $\mu$ and $c$ known from the simulations.

\begin{figure}[!htb]
\centering{}\centering \includegraphics[width=0.47\linewidth]{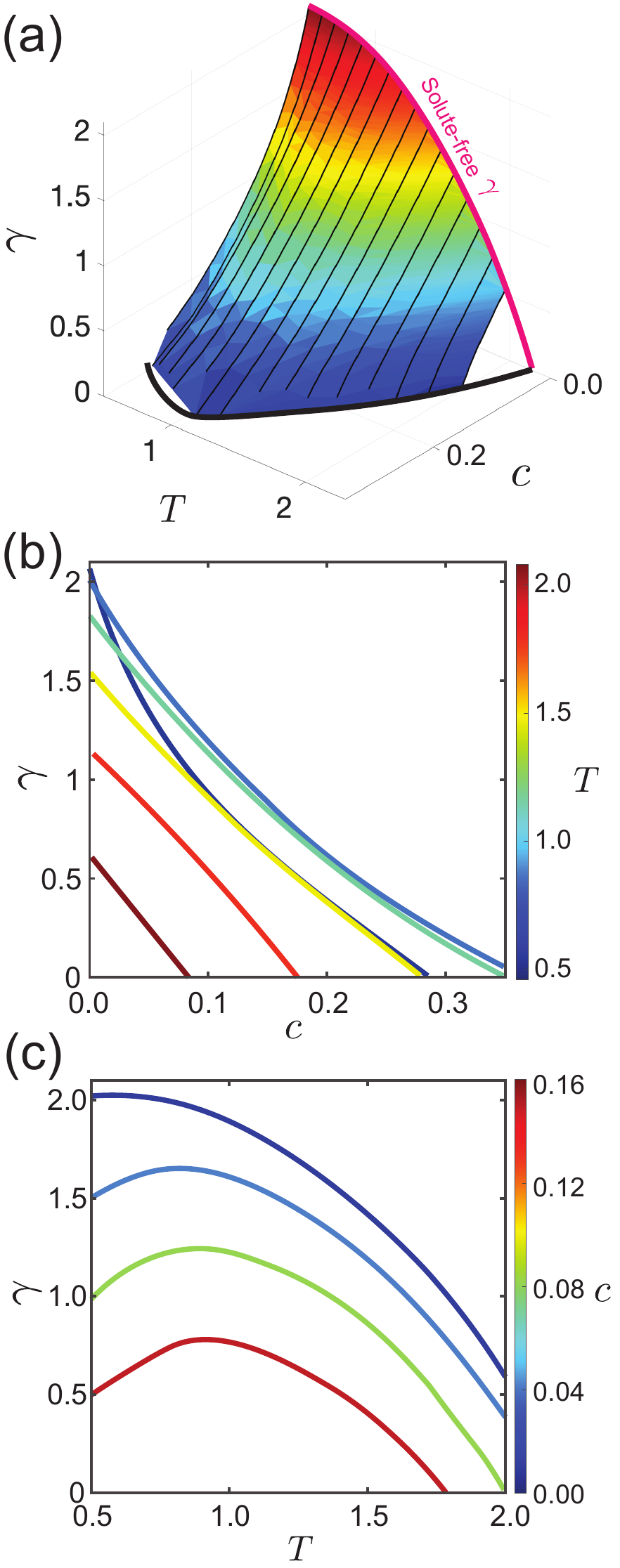}
\caption{(a) Three-dimensional surface plot of GB free energy $\gamma$ as
a function of temperature $T$ and solute concentration $c$. The
red and black curves show the results for the solute-free system and
the boundary of single-crystalline stability at which $\gamma=0$.
(b) Isothermal cross-sections of the surface in (a). (c) Iso-concentration
cross-sections of the surface in (a). The segregation energy is $J_{sg}=2$. }
\label{fig:InterfacialEnergy}
\end{figure}

Fig.~\ref{fig:InterfacialEnergy}(a) presents a three-dimensional
surface plot of the function $\gamma(T,c)$ computed for $J_{sg}=2$.
The $c=0$ cross-section of the surface is the solute-free GB free
energy $\gamma(T)$ reported in section \ref{subsec:gamma_pure}.
The surface crosses the $\gamma=0$ plane along the curve (shown in
black) that matches the $T$ versus $c$ phase diagram in Fig.~\ref{fig:CriticalTemperature}.
Recall that this curve represents the boundary of single-crystalline
stability. Beyond this boundary, single-crystalline alloys must transform
into a liquid-like state or a stable polycrystal. The observed close
agreement between the two independent calculations confirms the link
between the polycrystal stabilization and the $\gamma=0$ condition. 

The $\gamma$ versus $c$ plots in Fig.~\ref{fig:InterfacialEnergy}(b)
are isothermal cross-sections of the surface diagram in Fig.~\ref{fig:InterfacialEnergy}(a).
At a fixed temperature, the GB free energy monotonically decreases
with solute concentration, as predicted by the adsorption equation
in Eg.(\ref{eq:adsorption}) considering that $d\mu/dc>0$. Fig.~\ref{fig:InterfacialEnergy}(c)
shows the temperature dependencies of $\gamma$ at fixed alloy compositions.
The plots reveal a competition between the GB disordering and GB segregation.
In dilute alloys, $\gamma$ monotonically decreases with temperature
due to the structural disordering of GBs at high temperatures, which
increases the GB excess entropy. This trend is typical of elemental
GBs and is followed by the present model, see Fig.~\ref{fig:IsingModel_Interface}(a).
As the solute concentration increases, the curves turn over at low
temperatures. This reversal is due to the increase in GB segregation
at low temperatures. The segregation causes a decrease in $\gamma$,
which eventually overpowers the disordering effect.

\section{Solute segregation and grain boundary morphology\label{sec:Solute-segregation}}

In this section, we perform a more detailed investigation of GB segregation
and GB morphology as a function of alloy composition, temperature,
and the segregation energy $J_{sg}$. We use the KMC algorithm because
our goal is to study GBs in contact with nanoscale grains in a closed
system, in which the solute supply to the GBs is controlled by diffusion.
When we are only interested in equilibrium properties, we implement
fast diffusion conditions to accelerate the approach to equilibrium.
As before, we consider two GBs parallel to the $x$-axis with fully
periodic boundary conditions. The GBs are characterized by the excess
segregation $\tilde{N}$, intrinsic width $w_{I}$, and the mean-square
capillary width $w^{2}$. $\tilde{N}$ was computed as discussed in
section \ref{sec:TDI}, and the GB widths were calculated by the same
method as in the solute-free system (section \ref{subsec:widths}). 

\begin{figure}[!htb]
\centering{}\centering \includegraphics[width=0.38\linewidth]{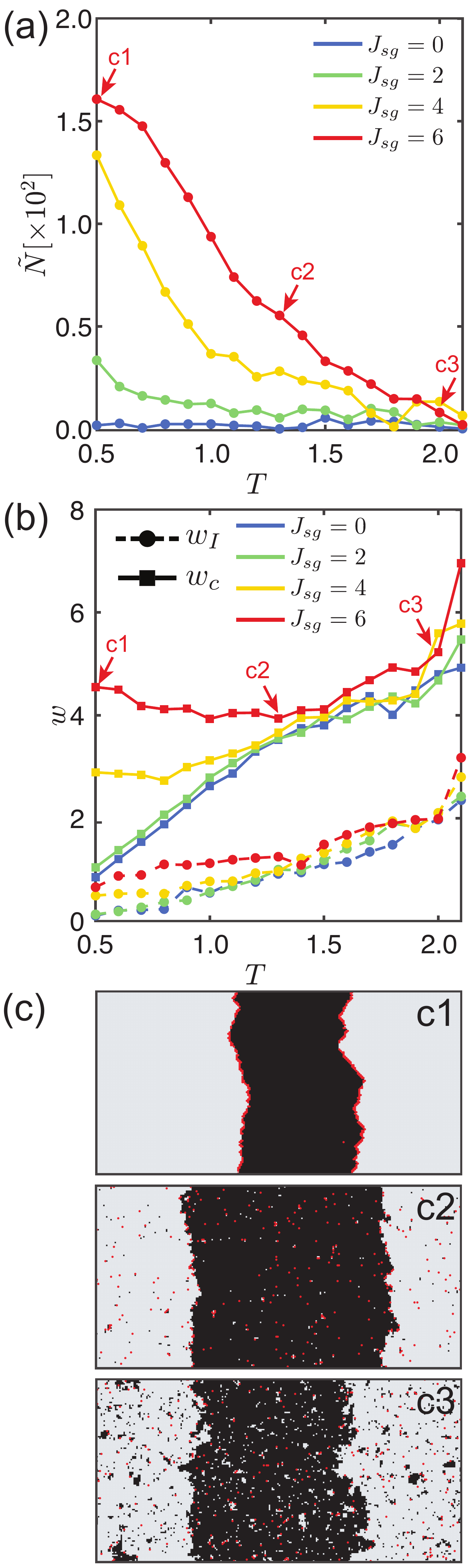}
\caption{GB segregation $\tilde{N}$ (a) and GB widths $w_{I}$ and $w_{c}$
(b) as a function of temperature for several $J_{sg}$ values. The
images in (c) show the structures at representative points c1, c2,
and c3. The red dots represent solute atoms. The average solute concentration
in the system is $c=0.01$.}
\label{fig:InterfacialCharacteristicForBetas}
\end{figure}

In Fig.~\ref{fig:InterfacialCharacteristicForBetas}(a), we plot
the equilibrium amount of solute segregation $\tilde{N}$ as a function
of temperature for the average solute concentration $c=0.01$. (The
results for higher concentrations are similar except that all $\tilde{N}$
values are larger.) The system size is $128\times256$, and the calculations
have been performed for several values of $J_{sg}$. Each curve ends
when the GB disintegrates transforming into a liquid-like state. The
plots show that the segregation is high at low temperatures and decreases
with temperature. This trend is typical for alloys with a positive
segregation energy. As $J_{sg}$ increases, the curves shift upward
due to the stronger segregation. 

In Fig.~\ref{fig:InterfacialCharacteristicForBetas}(b), we plot
the equilibrium intrinsic GB width $w_{I}$ and the capillary width
$w_{c}\equiv\sqrt{w^{2}}$ as a function of temperature for several
$J_{sg}$ values. The plots show that both widths increase with temperature
due to the thermal fluctuations and partial GB disordering. In the
alloy, both widths are systematically larger that in the solute-free
system ($J_{sg}=0$). At a fixed temperature, both widths increase
with $J_{sg}$. In other words, segregation thickness GBs and increases
the amplitude of capillary waves. Fig.~\ref{fig:InterfacialCharacteristicForBetas}(c)
shows the GB morphologies at three representative points labeled c1,
c2, and c3. At the lowest-temperature point c1, the GBs are nearly
saturated with the solute (explaining the large $\tilde{N}$ value)
and display a high degree of unevenness with some amount of faceting.
At a lower temperature (point c2), more solute is distributed inside
the grains at the expense of GBs (smaller $\tilde{N}$), and the faceting
nearly vanishes. At the high temperature approaching the bulk critical
point (c3), the solute atoms are scattered nearly uniformly. The grains
are significantly disordered and contain a distribution of small clusters
of the opposite orientation. 

Note that at $J_{sg}\geq4$, the capillary width exhibits a minimum
and increases with decreasing temperature at low temperatures. This
behavior can be attributed to a decrease in the GB stiffness caused
by the solute segregation. Recall that in section \ref{subsec:stiffness},
we were able to extract the GB stiffness of solute-free GBs from the
capillary fluctuation amplitudes. It would be tempting to apply the
capillary fluctuation method to quantify the GB stiffness in the alloys.
However, as discussed in Refs.~\citep{Mishin:2023aa,Mishin:2023ab},
Eqs.(\ref{eq:CFM}) and (\ref{eq:3.12-1}) cannot be applied to alloys,
except in the limiting cases of exceedingly fast or exceedingly slow
solute diffusion. Nevertheless, the predicted inverse correlation
between the capillary wave amplitudes and GB stiffness remains qualitatively
valid. We can, thus, infer from Fig.~\ref{fig:InterfacialCharacteristicForBetas}(b)
that GB segregation reduces the GB stiffness, and that the extent
of this reduction increases with decreasing temperature. 

\begin{figure}[!htb]
\centering \includegraphics[width=0.6\linewidth]{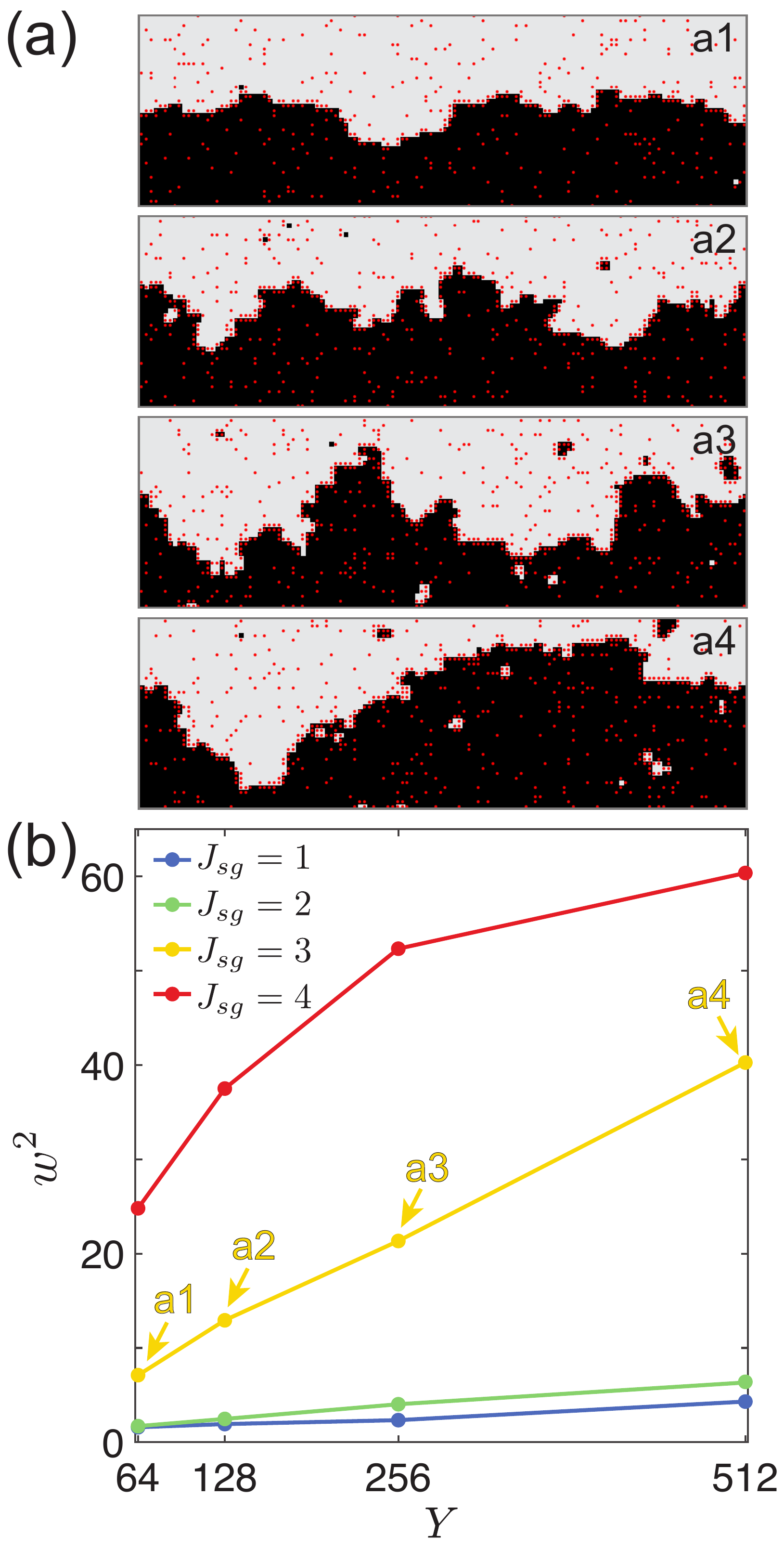}
\caption{(a) Representative snapshots showing GB morphology for four system
sizes $Y$. The red points represent solute atoms. The simulation
temperature is $T=0.7$, the average solute concentration is $c=0.1$,
and the GB segregation energy is $J_{sg}=3$. (b) Capillary width
as a function of system size for four values of $J_{sg}$. }
\label{fig:AspectRatio}
\end{figure}

To further demonstrate the link between GB segregation and capillary
fluctuations, we studied a set of closed systems with the same size
$X=128$ parallel to the GBs and four different sizes $Y=64$, $128$,
$256$, and $512$ in the normal direction. Initially, the solute
atoms were distributed uniformly across the system with a concentration
of $c=0.1$. KMC simulations were run at the temperature of $T=0.7$.
During the simulations, the atoms diffused towards the GBs and formed
segregation atmospheres. Since the total amount of solute in the system
increased with the size $Y$, more solute was available to form the
GB segregation. The limit of $Y\rightarrow\infty$ would create the
open-system condition, but for the finite $Y$ values tested here,
the system remained closed. The GB images in Figure~\ref{fig:AspectRatio}(a)
show that the amount of segregation and the capillary wave amplitude
both increase as the grain size increases. This correlation is quantified
in Figure~\ref{fig:AspectRatio}(b) showing that the capillary GB
width increases monotonically with the system size. The rate of increase
depends on the segregation energy $J_{sg}$, with higher $J_{sg}$
values leading to larger capillary wave amplitudes for the same system
size.

\begin{figure}
\begin{centering}
\includegraphics[width=1\textwidth]{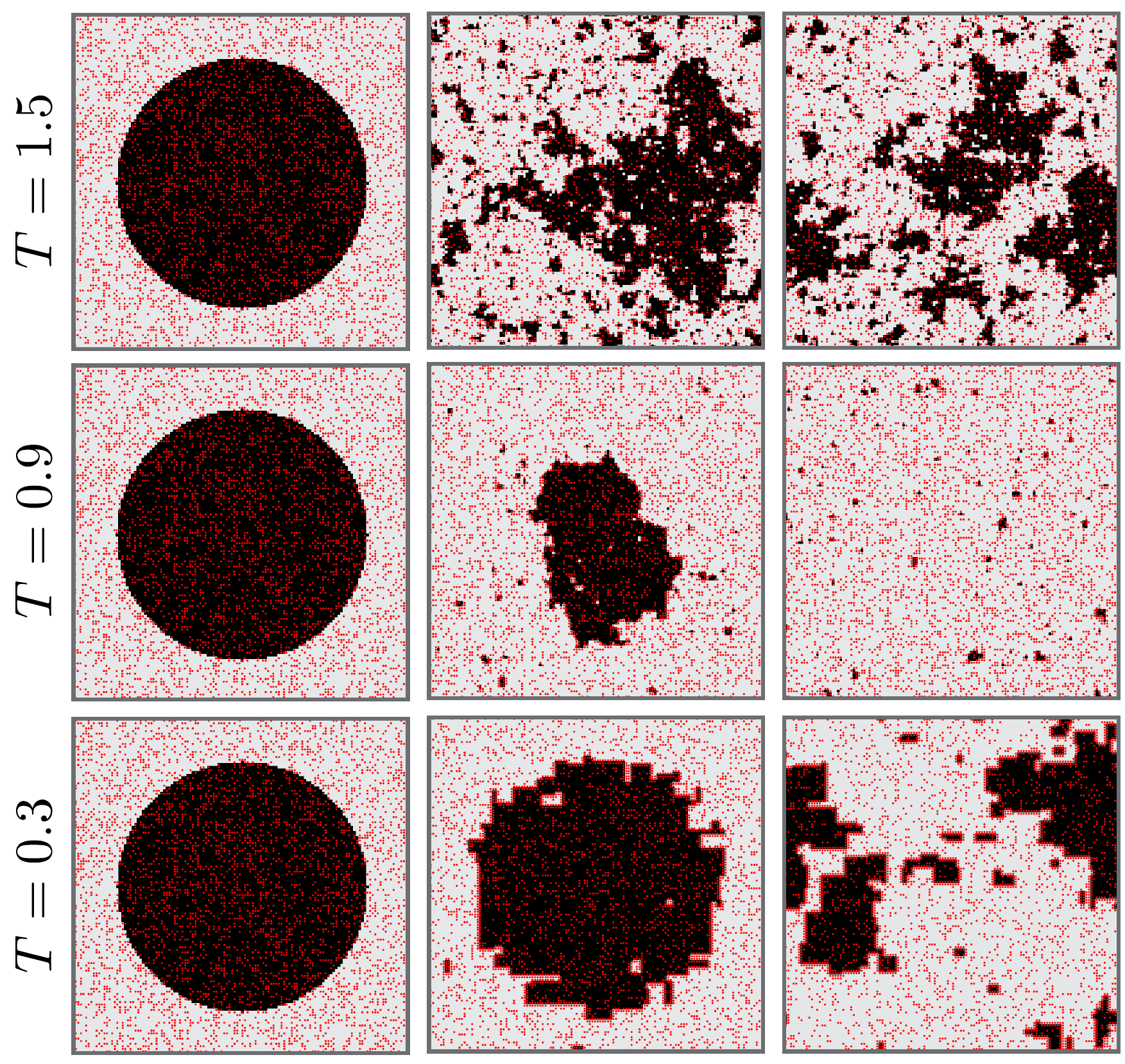}
\par\end{centering}
\caption{Single grain evolution in an alloy with $c=0.25$ at three different
temperatures. The left, middle, and right columns represent the initial,
intermediate, and fully equilibrated structures, respectively. The
results were obtained by KMC simulations with $J_{sg}=2$.\label{fig:grain-evolution}}
\end{figure}

\begin{figure}
\begin{centering}
\includegraphics[width=0.51\textwidth]{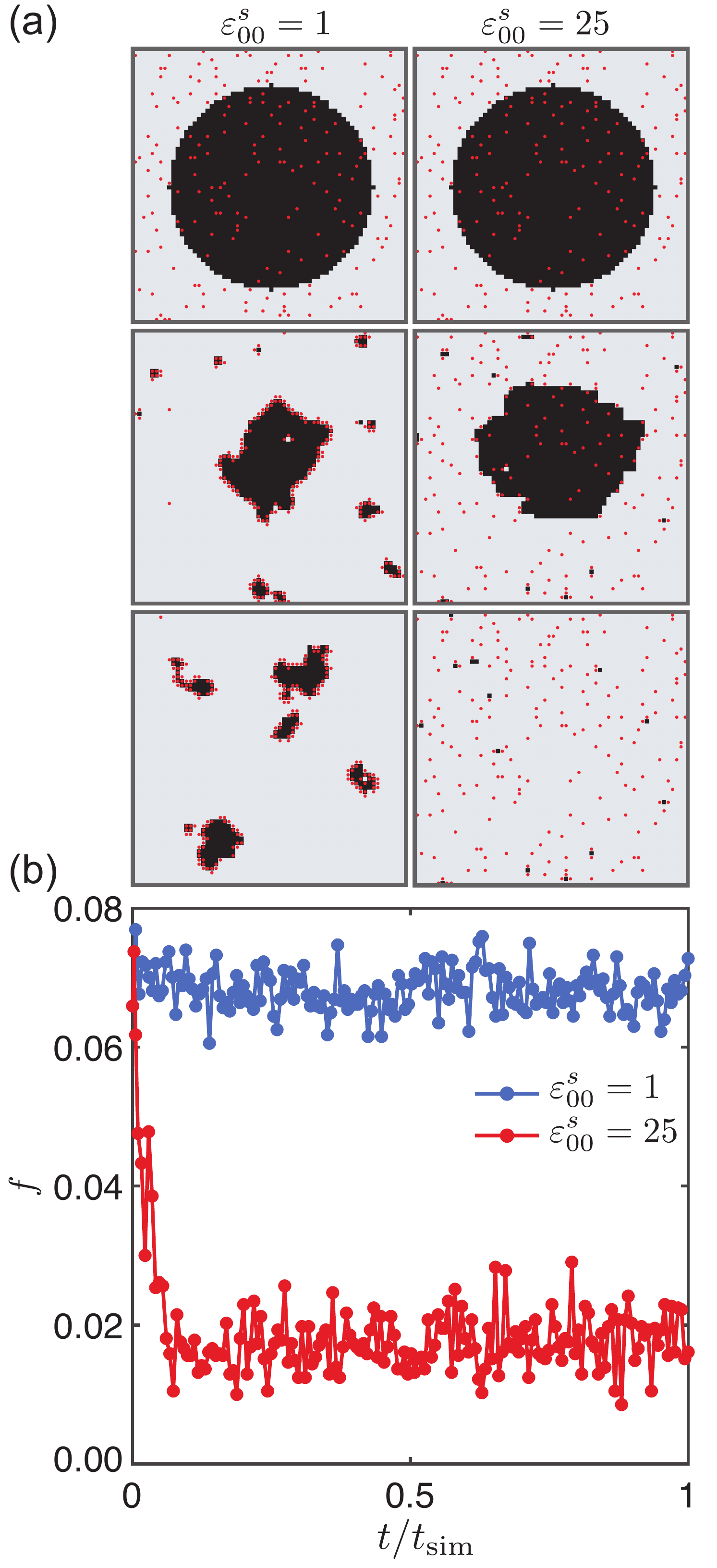}
\par\end{centering}
\caption{(a) Demonstration of the solute diffusivity effect on grain evolution.
The circular grain with the initially random distribution of the solute
(top row) evolves into different states (bottom row), depending on
the solute diffusion rate. Left column: fast diffusion ($\varepsilon_{00}^{s}=1$).
The grain breaks into smaller grains and forms a stable polycrystal.
Right column: slow diffusion ($\varepsilon_{00}^{s}=25$). The grain
shrink and disappears. (b) The fraction of GB sites $f$ as a function
of time $t$ normalized to the total simulation time $t_{\mathrm{sim}}$.\label{fig:(a)-solute-diffusion}}

\end{figure}

Finally, we return to the role of finite solute diffusion rates in
the full stabilization. To demonstrate this role, we have created
a circular grain similar to the one discussed previously (cf.~Fig.~\ref{fig:grain-evolution}),
except that the simulation parameters are now $T=0.7$, $c=0.05$,
and $J_{sg}=7$. This situation represents a GB that broke away from
its segregation atmosphere and found itself inside a grain. Two diffusion
regimes will be considered. First, when the solute diffusion barrier
is low ($\varepsilon_{00}^{s}=1$) and thus diffusion is fast, the
solute quickly catches up with the GB's new position and forms a new
segregation atmosphere (Fig.~\ref{fig:(a)-solute-diffusion}). The
GB free energy falls below zero and the grain starts increasing its
area by breaking into smaller grains. The grain refinement process
continues until the $\gamma=0$ condition is met and the system reaches
equilibrium. This process is illustrated in Fig.~\ref{fig:(a)-solute-diffusion}(a).
The fraction of GB sites $f$ is plotted as a function of time in
Fig.~\ref{fig:(a)-solute-diffusion}(b), demonstrating that the system
has reached equilibrium. This scenario is consistent with the previous
theoretical analyses and Metropolis Monte Carlo simulations effectively
implementing the fast diffusion conditions. In the the second regime,
the diffusion barrier is higher ($\varepsilon_{00}^{s}=25$) and the
solute diffusion is much slower. Since the initial GB free energy
is positive, the grain shrinks and soon disappears on a time scale
much shorter then the time scale of diffusion. Fig.~\ref{fig:(a)-solute-diffusion}(b)
shows that $f$ quickly drops to the background level ($f$ includes
not only the actual GBs but also the clusters of misoriented cells
inside the grains, which the algorithm treats as tiny ``grains'').
Once the grain is gone, it is highly unlikely that new grains can
nucleate homogeneously even if the simulation is run long enough to
allow significant diffusion. The system will remain single-crystalline
and with a higher free energy than the fully equilibrated polycrystal.
This example shows that changes in the solute diffusion rate can lead
to different stable or metastable states starting from the same initial
structure, chemical composition, and temperature.

\section{Concluding remarks}

The model proposed here describes GB migration coupled to solute diffusion.
The grain orientations are represented by a 2D Ising model, and the
solute atoms diffuse by nearest-neighbor jumps with environmentally-dependent
barriers. Despite its simplicity, the model captures a variety of
physical effects, including GB disordering at high temperatures, GB
segregation of the solute atoms, solute-solute interactions (which
can be different in GBs and inside the grains), and accelerated solute
diffusion along GBs relative to the grains. One of the limitations
of the model is that it only represents two possible lattice orientations.
As a result, polycrystalline structures are composed of hierarchically
enclosed grains, with one enclosed grain enclosing another, and so
on. As such, the model cannot represent triple junctions of grains,
which are typical structural elements in polycrystalline materials.
In addition, the underlying square lattice creates strongly anisotropic
GB properties and leads to GB faceting at low temperatures. In the
future, the model can be extended to higher-symmetry lattices (such
as the hexagonal lattice) to reduce the anisotropy and include several
lattice orientations using a Potts model. Extension to 3D polycrystals
should also be possible. 

Based on this model, we developed a rejection-free KMC algorithm implementing
the two concurrent processes: lattice reorientations and solute atom
jumps. This KMC algorithm distinguishes our work from previous simulations,
which implemented an Ising or Potts model with solutes using a Metropolis
Monte Carlo algorithm with unconditional acceptance of trial moves
when $E_{ij}\leq0$ \citep{Mendelev:2002wv,Mendelev:2001wk,Mendelev:2001aa,Liu:1998ab,Liu:1999aa}.
In addition to simulating unphysical dynamics, many implementations
of the Metropolis algorithm had a global character. The swaps of randomly
selected pairs of atoms, which can be far apart, can effectively sample
the configuration space of an equilibrium system or rapidly drive
the system towards equilibrium. However, non-equilibrium processes,
such as diffusion-controlled redistribution of solute atoms, occur
unrealistically fast. The proposed KMC algorithm correctly represents
solute diffusion as a random walk by nearest-neighbor jumps with adjustable
jump barriers. Such barriers are coupled to the lattice orientations,
allowing us to control diffusion rates in the perfect lattice and
in defected regions such as GBs.

The model reproduces the fully stabilized polycrystalline state at
relatively low temperatures in a certain range of segregation energies
and solute concentrations. If solute diffusion is sufficiently fast,
the alloy transforms into a fully stabilized polycrystal by breaking
the initial large grain into fragments, each representing a new grain.
Alternatively, an initially planar GB can develop large capillary
waves before breaking into smaller grains. In reality, there could
be other grain refinement mechanisms not captured by this model. 

The solute diffusion rate plays a significant role in the formation
of fully stabilized structures. The present model suggests that slow
solute diffusion can reroute the system evolution into non-equilibrium
or metastable states, and might possibly prevent the full stabilization.
The question of whether the full stabilization observed (under certain
conditions) in this model will be confirmed by more realistic models
with multiple lattice orientations is the left for future research.

\section*{Appendix: derivation of the $\gamma=0$ condition}

In this Appendix, we provide a rigorous derivation of the $\gamma=0$
condition of full stability of a polycrystalline alloy. Although the
derivation is based on Gibbs' interface thermodynamics, Gibbs treated
bulk phases as infinite reservoirs of the chemical components and
heat. Here, we apply Gibbs' method to a finite-size system. Also,
while Gibbs derived all equations starting from the fundamental equation
in the energy-entropy representation, we simplify the derivation by
using the free energy. This implies that we assume uniform temperature
across the system.

Consider an $m$-component polycrystalline alloy in internal thermodynamic
equilibrium, thermal equilibrium with a thermostat at a temperature
$T$, and mechanical equilibrium with a barostat at a pressure $p$.
The system is closed (no matter exchange with the environment) and
contains $N_{i}$ ($i=1,2,...,m$) atoms of each chemical species.
The system volume is $V$.

We assume that intensive properties of the solid solution inside the
grains are uniform and the same in all grains. We further adopt the
uniform boundary model, in which all grain boundaries (GBs) are characterized
by the same properties. Following Gibbs' interface thermodynamics
\citep{Gibbs}, we mentally partition the polycrystal into a grain
system and a GB system. The grain system is an imaginary uniform solid
solution filling the volume $V$ and having the same intensive properties
as the grain interiors. For example, if the number density of atoms
$i$ inside the grains is $n_{i}$, then the grain system contains
a total of $N_{i}^{g}=n_{i}V$ atoms of the component $i$. 

The fundamental equation describing the grain system is its free energy
$F^{g}$ expressed as a function of $T$, $V$, and the numbers of
atoms $N_{i}^{g}$:
\begin{equation}
F^{g}=F^{g}\left(T,V,N_{1}^{g},N_{2}^{g},...,N_{m}^{g}\right).\label{eq:F_g}
\end{equation}
The remaining free energy, $\tilde{F}$, is attributed to the GB system
and is called the total GB excess free energy. Thus, the total free
energy of the polycrystal, $F$, is partitioned into $F^{g}$ and
$\tilde{F}$:
\begin{equation}
F=F^{g}+\tilde{F}.\label{eq:F_tot}
\end{equation}
Similarly, we introduce excesses of all other extensive properties.
For example, the excess amounts of the chemical components are $\tilde{N}_{i}=N_{i}-N_{i}^{g}$,
the excess GB entropy is $\tilde{S}=S-S^{g}$ (where $S^{g}=-\partial F^{g}/\partial T$),
etc. 

The central postulate of the interface thermodynamics is that the
interfaces (in this case, GBs) follow their own fundamental equation
in the form
\begin{equation}
\tilde{F}=\tilde{F}\left(T,A,\tilde{N}_{1},\tilde{N}_{2},...,\tilde{N}_{m}\right),\label{eq:F_gb}
\end{equation}
where $A$ is the total GB area. In other words, the GB system can
be treated as 2D phase on par with the 3D (bulk) phase forming the
grains \citep{Frolov:2015ab}. Although the two phases have a different
dimensionality, they are described by mathematically similar fundamental
equations. The only difference is that the volume $V$ of the bulk
phase is replaced by the area $A$ (2D ``volume'') of GBs. It follows
that all thermodynamic equations describing a bulk phase can be immediately
rewritten for the GB phase by merely changing the notations.

The grain free energy $F^{g}$ in Eq.(\ref{eq:F_g}) is a homogeneous
first degree function of the variables $V$ and all $N_{i}^{g}$.
By Euler's theorem,
\begin{equation}
F^{g}=-pV+\sum_{i}\mu_{i}N_{i}^{g},\label{eq:F_g_1}
\end{equation}
where $\mu_{i}=\partial F^{g}/\partial N_{i}^{g}$ are the chemical
potentials in the grains and we used the mechanical equilibrium condition
\begin{equation}
p=-\dfrac{\partial F^{g}}{\partial V}.\label{eq:p}
\end{equation}
Taking the differential of $F^{g}$, we have
\begin{equation}
dF^{g}=-S^{g}dT-pdV+\sum_{i}\mu_{i}dN_{i}^{g}.\label{eq:dF_g}
\end{equation}
Combining Eqs.(\ref{eq:F_g_1}) and (\ref{eq:dF_g}), we obtain the
Gibbs-Duhem equation
\begin{equation}
-S^{g}dT+Vdp-\sum_{i}N_{i}^{g}d\mu_{i}=0.\label{eq:dF_g-1}
\end{equation}
We can now write down similar equations for the GB system:
\begin{equation}
\tilde{F}=\dfrac{\partial\tilde{F}}{\partial A}A+\sum_{i}\dfrac{\partial\tilde{F}}{\partial\tilde{N}_{i}}\tilde{N}_{i},\label{eq:F_g_1-1}
\end{equation}
\begin{equation}
d\tilde{F}=\dfrac{\partial\tilde{F}}{\partial T}dT+\dfrac{\partial\tilde{F}}{\partial A}dA+\sum_{i}\dfrac{\partial\tilde{F}}{\partial\tilde{N}_{i}}d\tilde{N}_{i},\label{eq:F_g_1-1-1}
\end{equation}
\begin{equation}
\dfrac{\partial\tilde{F}}{\partial T}dT-Ad\left(\dfrac{\partial\tilde{F}}{\partial A}\right)-\sum_{i}\tilde{N}_{i}d\left(\dfrac{\partial\tilde{F}}{\partial\tilde{N}_{i}}\right)=0.\label{eq:dF_g-1-1}
\end{equation}
In a closed system, the mass conservation constraint dictates $d\tilde{N}_{i}+dN_{i}^{g}=0$
for every species $i$. Also, we define
\begin{equation}
\gamma=\dfrac{\partial\tilde{F}}{\partial A}\label{eq:gamma_1}
\end{equation}
and call it the GB free energy or GB tension. This quantity is a 2D
analog of the 3D pressure $p$ (compare with Eq.(\ref{eq:p})). As
a result, the above equations take the form
\begin{equation}
\tilde{F}=\gamma A+\sum_{i}\dfrac{\partial\tilde{F}}{\partial\tilde{N}_{i}}\tilde{N}_{i},\label{eq:F_g_1-1-2}
\end{equation}
\begin{equation}
d\tilde{F}=-\tilde{S}dT+\gamma dA-\sum_{i}\dfrac{\partial\tilde{F}}{\partial\tilde{N}_{i}}dN_{i}^{g},\label{eq:F_g_1-1-1-1}
\end{equation}
\begin{equation}
-\tilde{S}dT-Ad\gamma-\sum_{i}\tilde{N}_{i}d\left(\dfrac{\partial\tilde{F}}{\partial\tilde{N}_{i}}\right)=0.\label{eq:dF_g-1-1-1}
\end{equation}

Next, we will need the total free energy $F$ and its differential
$dF$. These are readily obtained by adding the respective equations
from above:
\begin{equation}
F=F^{g}+\tilde{F}=-pV+\gamma A+\sum_{i}\mu_{i}N_{i}^{g}+\sum_{i}\dfrac{\partial\tilde{F}}{\partial\tilde{N}_{i}}\tilde{N}_{i},\label{eq:F1}
\end{equation}
\begin{equation}
dF=dF^{g}+d\tilde{F}=-SdT-pdV+\gamma dA+\sum_{i}\left(\mu_{i}-\dfrac{\partial\tilde{F}}{\partial\tilde{N}_{i}}\right)dN_{i}^{g}.\label{eq:dF1}
\end{equation}
We can now formulate the conditions of chemical equilibrium between
the GBs and the grains. Suppose $T$, $V$, and $A$ are fixed, and
an infinitesimal amount $dN_{i}^{g}$ of atoms $i$ moves from the
GBs into the grains. The first-order variation of the total free energy
with respect to this variation must be zero. It follows that the last
term in Eq.(\ref{eq:dF1}) must be identically zero. Thus, the GB-grain
equilibrium condition is
\begin{equation}
\dfrac{\partial\tilde{F}}{\partial\tilde{N}_{i}}=\mu_{i}\label{eq:mu}
\end{equation}
for every species $i$. Inserting this condition into Eqs.(\ref{eq:F1})
and (\ref{eq:dF1}), we finally obtain
\begin{equation}
F=-pV+\gamma A+\sum_{i}\mu_{i}N_{i},\label{eq:F1-1}
\end{equation}
\begin{equation}
dF=-SdT-pdV+\gamma dA.\label{eq:dF1-1}
\end{equation}
We emphasize that Eq.(\ref{eq:dF1-1}) represents equilibrium variations
maintaining the GB-grain equilibrium.

\begin{figure}
\begin{centering}
\includegraphics[width=0.45\textwidth]{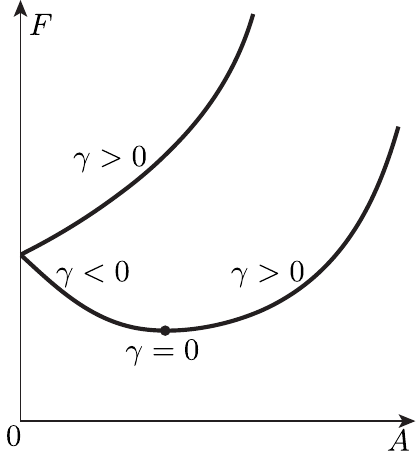}
\par\end{centering}
\caption{Schematic presentation of the free energy energy $F$ as a function
of grain boundary area $A$ in a polycrystalline material. Upper curve:
commonly observed increase of free energy with grain refinement when
the GB free energy $\gamma$ is positive. Lower curve: hypothetical
decrease of free energy with grain refinement when $\gamma<0$, followed
by increase after $\gamma$ changes sign. The point with $\gamma=0$
represent a fully stable polycrystalline material.\label{fig:Schematic-F(A)}}

\end{figure}

Eq.(\ref{eq:dF1-1}) is what we wanted to derive. It shows that 
\begin{equation}
\left(\dfrac{\partial F}{\partial A}\right)_{T,V}=\gamma.\label{eq:gamma_2}
\end{equation}
If $\gamma>0$, the free energy increases with GB area. In other words,
to decrease the free energy, the grains must grow. If $\gamma<0$,
the polycrystal will tend to decrease the grain size (increase $A$).
To obtain a stable polycrystal with a finite grain size, $\gamma$
must change sign as shown schematically in Fig.~\ref{fig:Schematic-F(A)}.
The free energy minimum is reached at $\gamma=0$.

Although Eqs.(\ref{eq:F1-1}) and (\ref{eq:dF1-1}) look similar to
the known equations from Gibbs \citep{Gibbs}, they have been derived
for a finite-size closed system, and their meaning is subtly different.
To appreciate the difference, recall that the GB free energy is commonly
defined as the reversible work of creating a unit GB area in a closed
system. This definition implies GB creation in an initially single-crystalline
system having the same temperature and volume and containing the same
number of atoms of every species as the polycrystal. This definition
appears to be consistent with Eq.(\ref{eq:F1-1}), which can be rewritten
in the form
\begin{equation}
\gamma=\dfrac{F-\left(\sum_{i}\mu_{i}N_{i}-pV\right)}{A},\label{eq:gamma_3}
\end{equation}
where the term in the parentheses is the free energy of a uniform
solid solution. However, this is \emph{not} the free energy of the
imaginary solution that initially filled the volume $V$ and from
which the polycrystal has been formed. That solution follows the same
fundamental equation (\ref{eq:F_g}) as the grains but with a different
chemical composition:
\begin{equation}
F^{0}=F^{g}\left(T,V,N_{1},N_{2},...,N_{m}\right)=-p^{0}V+\sum_{i}\mu_{i}^{0}N_{i}.\label{eq:F_g-1}
\end{equation}
The chemical potentials in that solution, $\mu_{i}^{0}=\partial F^{g}/N_{i}$,
are different from those in the grains due to the GB segregation.
The pressure $p^{0}$ is generally also different. The free energy
change relative that original solution is
\begin{equation}
F-F^{0}=\gamma A-(p-p^{0})V+\sum_{i}\left(\mu_{i}-\mu_{i}^{0}\right)N_{i}.\label{eq:dF_4}
\end{equation}

Eq.(\ref{eq:dF_4}) shows that $\gamma A$ is generally different
from the free energy change $F-F^{0}$ relative to the original solution.
The difference arises because additional free energy is expended to
change the chemical composition of the uniform solid solution. The
GBs only know the chemical composition in the adjacent grains, and
this is what defines $\gamma$. The grain composition only represents
the composition of the original solution if the grains are large enough
or the system is open. In that case, $\mu_{i}\rightarrow\mu_{i}^{0}$,
$p\rightarrow p^{0}$, and we recover the known relation
\begin{equation}
\gamma=\dfrac{F-F^{0}}{A}.\label{eq:gamma_5}
\end{equation}


\begin{thebibliography}{50}
\expandafter\ifx\csname natexlab\endcsname\relax\def\natexlab#1{#1}\fi
\providecommand{\url}[1]{\texttt{#1}}
\providecommand{\href}[2]{#2}
\providecommand{\path}[1]{#1}
\providecommand{\DOIprefix}{doi:}
\providecommand{\ArXivprefix}{arXiv:}
\providecommand{\URLprefix}{URL: }
\providecommand{\Pubmedprefix}{pmid:}
\providecommand{\doi}[1]{\href{http://dx.doi.org/#1}{\path{#1}}}
\providecommand{\Pubmed}[1]{\href{pmid:#1}{\path{#1}}}
\providecommand{\bibinfo}[2]{#2}
\ifx\xfnm\relax \def\xfnm[#1]{\unskip,\space#1}\fi
\bibitem[{Weissm{\"u}ller(1994)}]{Weissmuller1994}
\bibinfo{author}{J.~Weissm{\"u}ller},
\newblock \bibinfo{title}{Alloy thermodynamics in nano-structures},
\newblock \bibinfo{journal}{J. Mater. Res.} \bibinfo{volume}{9}
  (\bibinfo{year}{1994}) \bibinfo{pages}{4--7}.
\bibitem[{Weissm{\"u}ller(1993)}]{Weissmuller:1993aa}
\bibinfo{author}{J.~Weissm{\"u}ller},
\newblock \bibinfo{title}{Alloy effects in nanostructures},
\newblock \bibinfo{journal}{Nanostructured Materials} \bibinfo{volume}{3}
  (\bibinfo{year}{1993}) \bibinfo{pages}{261--272}.
\bibitem[{Kirchheim(2002)}]{Kirchheim:2002aa}
\bibinfo{author}{R.~Kirchheim},
\newblock \bibinfo{title}{Grain coarsening inhibited by solute segregation},
\newblock \bibinfo{journal}{Acta Materialia} \bibinfo{volume}{50}
  (\bibinfo{year}{2002}) \bibinfo{pages}{413--419}.
\bibitem[{Liu and Kirchheim(2004)}]{Liu:2004aa}
\bibinfo{author}{F.~Liu}, \bibinfo{author}{R.~Kirchheim},
\newblock \bibinfo{title}{Nano-scale grain growth inhibited by reducing grain
  boundary energy through solute segregation},
\newblock \bibinfo{journal}{Journal of Crystal Growth} \bibinfo{volume}{264}
  (\bibinfo{year}{2004}) \bibinfo{pages}{385--391}.
\bibitem[{Kirchheim(2007{\natexlab{a}})}]{Kirchheim2007a}
\bibinfo{author}{R.~Kirchheim},
\newblock \bibinfo{title}{Reducing grain boundary, dislocation line and vacancy
  formation energies by solute segregation. {I.} {Theoretical} background},
\newblock \bibinfo{journal}{Acta Mater.} \bibinfo{volume}{55}
  (\bibinfo{year}{2007}{\natexlab{a}}) \bibinfo{pages}{5129--5138}.
\bibitem[{Kirchheim(2007{\natexlab{b}})}]{Kirchheim2007b}
\bibinfo{author}{R.~Kirchheim},
\newblock \bibinfo{title}{Reducing grain boundary, dislocation line and vacancy
  formation energies by solute segregation. {II.} {Experimental} evidence and
  consequences},
\newblock \bibinfo{journal}{Acta Mater.} \bibinfo{volume}{55}
  (\bibinfo{year}{2007}{\natexlab{b}}) \bibinfo{pages}{5139--5148}.
\bibitem[{Krill et~al.(2005)Krill, Ehrhardt, and Birringer}]{Krill:2005aa}
\bibinfo{author}{C.~E. Krill}, \bibinfo{author}{H.~Ehrhardt},
  \bibinfo{author}{R.~Birringer},
\newblock \bibinfo{title}{Thermodynamic stabilization of nanocrystallinity},
\newblock \bibinfo{journal}{International Journal of Materials Research}
  \bibinfo{volume}{96} (\bibinfo{year}{2005}) \bibinfo{pages}{1134--1141}.
\bibitem[{Schvindlerman and Gottstein(2006)}]{Schvindlerman2006}
\bibinfo{author}{L.~S. Schvindlerman}, \bibinfo{author}{G.~Gottstein},
\newblock \bibinfo{title}{Unexplored topics and potentials of grain boundary
  engineering},
\newblock \bibinfo{journal}{Scripta Mater.} \bibinfo{volume}{54}
  (\bibinfo{year}{2006}) \bibinfo{pages}{1041--1045}.
\bibitem[{Detor and Schuh(2007)}]{Detor2007}
\bibinfo{author}{A.~J. Detor}, \bibinfo{author}{C.~A. Schuh},
\newblock \bibinfo{title}{Grain boundary segregation, chemical ordering and
  stability of nanocrystalline alloys: Atomistic computer simulations in the
  ni-w system},
\newblock \bibinfo{journal}{Acta Mater.} \bibinfo{volume}{55}
  (\bibinfo{year}{2007}) \bibinfo{pages}{4221--4232}.
\bibitem[{Chookajorn et~al.(2012)Chookajorn, Murdoch, and
  Schuh}]{Chookajorn2012}
\bibinfo{author}{T.~Chookajorn}, \bibinfo{author}{H.~A. Murdoch},
  \bibinfo{author}{C.~A. Schuh},
\newblock \bibinfo{title}{Design of stable nano-crystalline alloys},
\newblock \bibinfo{journal}{Science} \bibinfo{volume}{337}
  (\bibinfo{year}{2012}) \bibinfo{pages}{951--953}.
\bibitem[{Chookajorn and Schuh(2014)}]{Chookajorn2014}
\bibinfo{author}{T.~Chookajorn}, \bibinfo{author}{C.~A. Schuh},
\newblock \bibinfo{title}{Thermodynamics of stable nanocrystalline alloys: A
  {Monte Carlo} analysis},
\newblock \bibinfo{journal}{Phys. Rev. {\rm B}} \bibinfo{volume}{89}
  (\bibinfo{year}{2014}) \bibinfo{pages}{064102}.
\bibitem[{Kalidindi and Schuh(2017)}]{Kalidindi:2017aa}
\bibinfo{author}{A.~R. Kalidindi}, \bibinfo{author}{C.~A. Schuh},
\newblock \bibinfo{title}{Stability criteria for nanocrystalline alloys},
\newblock \bibinfo{journal}{Acta Mater.} \bibinfo{volume}{132}
  (\bibinfo{year}{2017}) \bibinfo{pages}{128--137}.
\bibitem[{A.~R.~Kalidindi and Schuh(2015)}]{Kalidindi:2017bb}
\bibinfo{author}{T.~C. A.~R.~Kalidindi}, \bibinfo{author}{C.~A. Schuh},
\newblock \bibinfo{title}{Nanocrystalline materials at equilibrium: A
  thermodynamic review},
\newblock \bibinfo{journal}{JOM} \bibinfo{volume}{67} (\bibinfo{year}{2015})
  \bibinfo{pages}{2834--2843}.
\bibitem[{Kalidindi and Schuh(2017)}]{Kalidindi:2017cc}
\bibinfo{author}{A.~R. Kalidindi}, \bibinfo{author}{C.~A. Schuh},
\newblock \bibinfo{title}{Stability criteria for nanocrystalline alloys},
\newblock \bibinfo{journal}{J. Mater. Res.} \bibinfo{volume}{32}
  (\bibinfo{year}{2017}) \bibinfo{pages}{1093--2002}.
\bibitem[{Murdoch and Schuh(2013)}]{Murdoch:2013aa}
\bibinfo{author}{H.~A. Murdoch}, \bibinfo{author}{C.~A. Schuh},
\newblock \bibinfo{title}{Stability of binary nanocrystalline alloys against
  grain growth and phase separation},
\newblock \bibinfo{journal}{Acta Mater.} \bibinfo{volume}{61}
  (\bibinfo{year}{2013}) \bibinfo{pages}{2121--2132}.
\bibitem[{Trelewicz and Schuh(2009)}]{Trelewicz2009}
\bibinfo{author}{J.~R. Trelewicz}, \bibinfo{author}{C.~A. Schuh},
\newblock \bibinfo{title}{Grain boundary segregation and thermodynamically
  stable binary nanocrystalline alloys},
\newblock \bibinfo{journal}{Phys. Rev. {\rm B}} \bibinfo{volume}{79}
  (\bibinfo{year}{2009}) \bibinfo{pages}{094112}.
\bibitem[{Perrin and Schuh(2021)}]{Perrin-2021}
\bibinfo{author}{A.~E. Perrin}, \bibinfo{author}{C.~A. Schuh},
\newblock \bibinfo{title}{Stabilized nanocrystalline alloys: The intersection
  of grain boundary segregation with processing science},
\newblock \bibinfo{journal}{Annual Review of Materials Research}
  \bibinfo{volume}{51} (\bibinfo{year}{2021}) \bibinfo{pages}{241--268}.
\bibitem[{Koch et~al.(2008)Koch, Scattergood, Darling, and Semones}]{Koch2008}
\bibinfo{author}{C.~C. Koch}, \bibinfo{author}{R.~O. Scattergood},
  \bibinfo{author}{K.~A. Darling}, \bibinfo{author}{J.~E. Semones},
\newblock \bibinfo{title}{Stabilization of nanocrystalline grain sizes by
  solute additions},
\newblock \bibinfo{journal}{J. Mater. Sci.} \bibinfo{volume}{43}
  (\bibinfo{year}{2008}) \bibinfo{pages}{7264--7272}.
\bibitem[{Darling et~al.(2014)Darling, Tschopp, VanLeeuwen, Atwater, and
  Liu}]{Darling2014}
\bibinfo{author}{K.~A. Darling}, \bibinfo{author}{M.~A. Tschopp},
  \bibinfo{author}{B.~K. VanLeeuwen}, \bibinfo{author}{M.~A. Atwater},
  \bibinfo{author}{Z.~K. Liu},
\newblock \bibinfo{title}{Mitigating grain growth in binary nanocrystalline
  alloys through solute selection based on thermodynamic stability maps},
\newblock \bibinfo{journal}{Comp. Mater. Sci.} \bibinfo{volume}{84}
  (\bibinfo{year}{2014}) \bibinfo{pages}{255--266}.
\bibitem[{Saber et~al.(2013{\natexlab{a}})Saber, Kotan, Koch, and
  Scattergood}]{Saber2013}
\bibinfo{author}{M.~Saber}, \bibinfo{author}{H.~Kotan}, \bibinfo{author}{C.~C.
  Koch}, \bibinfo{author}{R.~O. Scattergood},
\newblock \bibinfo{title}{Thermodynamic stabilization of nanocrystalline binary
  alloys},
\newblock \bibinfo{journal}{J. Appl. Phys.} \bibinfo{volume}{113}
  (\bibinfo{year}{2013}{\natexlab{a}}) \bibinfo{pages}{065515}.
\bibitem[{Saber et~al.(2013{\natexlab{b}})Saber, Kotan, Koch, and
  Scattergood}]{Saber2013a}
\bibinfo{author}{M.~Saber}, \bibinfo{author}{H.~Kotan}, \bibinfo{author}{C.~C.
  Koch}, \bibinfo{author}{R.~O. Scattergood},
\newblock \bibinfo{title}{A predictive model for thermodynamic stability of
  grain size in nanocrystalline ternary alloys},
\newblock \bibinfo{journal}{J. Appl. Phys.} \bibinfo{volume}{114}
  (\bibinfo{year}{2013}{\natexlab{b}}) \bibinfo{pages}{103510}.
\bibitem[{Xing et~al.(2018)Xing, Kalidindi, Amram, and Schuh}]{Xing:2018aa}
\bibinfo{author}{W.~Xing}, \bibinfo{author}{A.~R. Kalidindi},
  \bibinfo{author}{D.~Amram}, \bibinfo{author}{C.~A. Schuh},
\newblock \bibinfo{title}{Solute interaction effects on grain boundary
  segregation in ternary alloys},
\newblock \bibinfo{journal}{Acta Materialia} \bibinfo{volume}{161}
  (\bibinfo{year}{2018}) \bibinfo{pages}{285--294}.
\bibitem[{Zhou and Luo(2014)}]{Zhou:2014aa}
\bibinfo{author}{N.~Zhou}, \bibinfo{author}{J.~Luo},
\newblock \bibinfo{title}{Developing thermodynamic stability diagrams for
  equilibrium-grain-size binary alloys},
\newblock \bibinfo{journal}{Materials Letters} \bibinfo{volume}{115}
  (\bibinfo{year}{2014}) \bibinfo{pages}{268--271}.
\bibitem[{Kalidindi et~al.(2015)Kalidindi, Chookajorn, and
  Schuh}]{Kalidindi:2015aa}
\bibinfo{author}{A.~R. Kalidindi}, \bibinfo{author}{T.~Chookajorn},
  \bibinfo{author}{C.~A. Schuh},
\newblock \bibinfo{title}{Nanocrystalline materials at equilibrium: A
  thermodynamic review},
\newblock \bibinfo{journal}{JOM} \bibinfo{volume}{67} (\bibinfo{year}{2015})
  \bibinfo{pages}{2834--2843}.
\bibitem[{Mendelev et~al.(2001)Mendelev, Srolovitz, and E}]{Mendelev:2001aa}
\bibinfo{author}{M.~I. Mendelev}, \bibinfo{author}{D.~J. Srolovitz},
  \bibinfo{author}{W.~E},
\newblock \bibinfo{title}{Grain-boundary migration in the presence of diffusing
  impurities: simulations and analytical models},
\newblock \bibinfo{journal}{Philos. Mag.} \bibinfo{volume}{81}
  (\bibinfo{year}{2001}) \bibinfo{pages}{2243--2269}.
\bibitem[{Mendelev and Srolovitz(2001)}]{Mendelev:2001wk}
\bibinfo{author}{M.~I. Mendelev}, \bibinfo{author}{D.~J. Srolovitz},
\newblock \bibinfo{title}{Kink model for extended defect migration in the
  presence of diffusing impurities: theory and simulation},
\newblock \bibinfo{journal}{Acta Materialia} \bibinfo{volume}{49}
  (\bibinfo{year}{2001}) \bibinfo{pages}{2843--2852}.
\bibitem[{Liu(1998)}]{Liu:1998ab}
\bibinfo{author}{J.~M. Liu},
\newblock \bibinfo{title}{Monte carlo simulation of solute aggregation on
  domain boundaries in binary alloys: Domain-boundary segregation and domain
  growth},
\newblock \bibinfo{journal}{Physical Review B} \bibinfo{volume}{58}
  (\bibinfo{year}{1998}) \bibinfo{pages}{633--640}.
\bibitem[{Liu et~al.(1999)Liu, Lim, and Liu}]{Liu:1999aa}
\bibinfo{author}{J.~M. Liu}, \bibinfo{author}{L.~C. Lim},
  \bibinfo{author}{Z.~G. Liu},
\newblock \bibinfo{title}{Monte carlo simulation of solute aggregation in
  binary alloys: Domain boundary precipitation and domain growth},
\newblock \bibinfo{journal}{Physical Review B} \bibinfo{volume}{60}
  (\bibinfo{year}{1999}) \bibinfo{pages}{7113--7126}.
\bibitem[{Vineyard(1957)}]{Vineyard:1957vo}
\bibinfo{author}{G.~H. Vineyard},
\newblock \bibinfo{title}{Frequency factors and isotope effects in solid state
  rate processes},
\newblock \bibinfo{journal}{Journal of Physics and Chemistry of Solids}
  \bibinfo{volume}{3} (\bibinfo{year}{1957}) \bibinfo{pages}{121--127}.
\bibitem[{Mishin(2023{\natexlab{a}})}]{Mishin:2023ab}
\bibinfo{author}{Y.~Mishin},
\newblock \bibinfo{title}{Stochastic model and kinetic monte carlo simulation
  of solute interactions with stationary and moving grain boundaries. ii.
  application to two-dimensional systems},
\newblock \bibinfo{journal}{Physical Review Materials} \bibinfo{volume}{7}
  (\bibinfo{year}{2023}{\natexlab{a}}) \bibinfo{pages}{063404}.
\bibitem[{Mishin(2023{\natexlab{b}})}]{Mishin:2023aa}
\bibinfo{author}{Y.~Mishin},
\newblock \bibinfo{title}{Stochastic model and kinetic monte carlo simulation
  of solute interactions with stationary and moving grain boundaries. i. model
  formulation and application to one-dimensional systems},
\newblock \bibinfo{journal}{Physical Review Materials} \bibinfo{volume}{7}
  (\bibinfo{year}{2023}{\natexlab{b}}) \bibinfo{pages}{063403}.
\bibitem[{Kaur et~al.(1995)Kaur, Mishin, and Gust}]{Kaur95}
\bibinfo{author}{I.~Kaur}, \bibinfo{author}{Y.~Mishin},
  \bibinfo{author}{W.~Gust}, \bibinfo{title}{Fundamentals of Grain and
  Interphase Boundary Diffusion}, \bibinfo{publisher}{Wiley},
  \bibinfo{address}{Chichester, West Sussex}, \bibinfo{year}{1995}.
\bibitem[{Mishin et~al.(1997)Mishin, Herzig, Bernardini, and Gust}]{Mishin97e}
\bibinfo{author}{Y.~Mishin}, \bibinfo{author}{C.~Herzig},
  \bibinfo{author}{J.~Bernardini}, \bibinfo{author}{W.~Gust},
\newblock \bibinfo{title}{Grain boundary diffusion: fundamentals to recent
  developments},
\newblock \bibinfo{journal}{Int. Mater. Reviews} \bibinfo{volume}{42}
  (\bibinfo{year}{1997}) \bibinfo{pages}{155}.
\bibitem[{Mishin and Herzig(1999)}]{Mishin99f}
\bibinfo{author}{Y.~Mishin}, \bibinfo{author}{C.~Herzig},
\newblock \bibinfo{title}{Grain boundary diffusion: recent progress and future
  research},
\newblock \bibinfo{journal}{Mater. Sci. Eng. {\rm A}} \bibinfo{volume}{260}
  (\bibinfo{year}{1999}) \bibinfo{pages}{55--71}.
\bibitem[{Bortz et~al.(1975)Bortz, Kalos, and Lebowitz}]{bortz1975new}
\bibinfo{author}{A.~B. Bortz}, \bibinfo{author}{M.~H. Kalos},
  \bibinfo{author}{J.~L. Lebowitz},
\newblock \bibinfo{title}{A new algorithm for monte carlo simulation of ising
  spin systems},
\newblock \bibinfo{journal}{Journal of Computational Physics}
  \bibinfo{volume}{17} (\bibinfo{year}{1975}) \bibinfo{pages}{10--18}.
\bibitem[{Landau and Lifshitz(2000)}]{Landau-Lifshitz-Stat-phys}
\bibinfo{author}{L.~D. Landau}, \bibinfo{author}{E.~M. Lifshitz},
  \bibinfo{title}{Statistical Physics, Part I}, volume~\bibinfo{volume}{5} of
  \textit{\bibinfo{series}{Course of Theoretical Physics}},
  \bibinfo{edition}{third} ed., \bibinfo{publisher}{Butterworth-Heinemann},
  \bibinfo{address}{Oxford}, \bibinfo{year}{2000}.
\bibitem[{Mishin(2015)}]{Mishin:2015ab}
\bibinfo{author}{Y.~Mishin},
\newblock \bibinfo{title}{Thermodynamic theory of equilibrium fluctuations},
\newblock \bibinfo{journal}{Annals of Physics} \bibinfo{volume}{363}
  (\bibinfo{year}{2015}) \bibinfo{pages}{48--97}.
\bibitem[{Onsager(1944)}]{Onsager:1944aa}
\bibinfo{author}{L.~Onsager},
\newblock \bibinfo{title}{Crystal statistics. i. a two-dimensional model with
  an order-disorder transition},
\newblock \bibinfo{journal}{Physical Review} \bibinfo{volume}{65}
  (\bibinfo{year}{1944}) \bibinfo{pages}{117--149}.
\bibitem[{Yang(1952)}]{Yang:1952aa}
\bibinfo{author}{C.~N. Yang},
\newblock \bibinfo{title}{The spontaneous magnetization of a two-dimensional
  ising model},
\newblock \bibinfo{journal}{Physical Review} \bibinfo{volume}{85}
  (\bibinfo{year}{1952}) \bibinfo{pages}{808--816}.
\bibitem[{Binder et~al.(2011)Binder, Block, Das, Virnau, and
  Winter}]{Binder2011}
\bibinfo{author}{K.~Binder}, \bibinfo{author}{B.~Block}, \bibinfo{author}{S.~K.
  Das}, \bibinfo{author}{P.~Virnau}, \bibinfo{author}{D.~Winter},
\newblock \bibinfo{title}{Monte carlo methods for estimating interfacial free
  energies and line tensions},
\newblock \bibinfo{journal}{J. Statist. Phys.} \bibinfo{volume}{144}
  (\bibinfo{year}{2011}) \bibinfo{pages}{690--729}.
\bibitem[{Hoyt et~al.(2001)Hoyt, Asta, and Karma}]{Hoyt01}
\bibinfo{author}{J.~J. Hoyt}, \bibinfo{author}{M.~Asta},
  \bibinfo{author}{A.~Karma},
\newblock \bibinfo{title}{Method for computing the anisotropy of the
  solid-liquid interfacial free energy},
\newblock \bibinfo{journal}{Phys. Rev. Lett.} \bibinfo{volume}{86}
  (\bibinfo{year}{2001}) \bibinfo{pages}{5530--5533}.
\bibitem[{Morris and Song(2002)}]{Morris02}
\bibinfo{author}{J.~Morris}, \bibinfo{author}{X.~Song},
\newblock \bibinfo{title}{The melting lines of model systems calculated from
  coexistence simulations},
\newblock \bibinfo{journal}{J. Chem. Phys.} \bibinfo{volume}{116}
  (\bibinfo{year}{2002}) \bibinfo{pages}{9352--9358}.
\bibitem[{Pun et~al.(2020)Pun, Yamakov, Hickman, Glaessgen, and
  Mishin}]{Pun:2020aa}
\bibinfo{author}{G.~P.~P. Pun}, \bibinfo{author}{V.~Yamakov},
  \bibinfo{author}{J.~Hickman}, \bibinfo{author}{E.~H. Glaessgen},
  \bibinfo{author}{Y.~Mishin},
\newblock \bibinfo{title}{Development of a general-purpose machine-learning
  interatomic potential for aluminum by the physically informed neural network
  method},
\newblock \bibinfo{journal}{Physical Review Materials} \bibinfo{volume}{4}
  (\bibinfo{year}{2020}) \bibinfo{pages}{113807}.
\bibitem[{Mishin(2014)}]{Mishin2014}
\bibinfo{author}{Y.~Mishin},
\newblock \bibinfo{title}{Calculation of the $\gamma/\gamma^\prime$ interface
  free energy in the {Ni-Al} system by the capillary fluctuation method},
\newblock \bibinfo{journal}{Modeling Simul. Mater. Sci. Eng.}
  \bibinfo{volume}{22} (\bibinfo{year}{2014}) \bibinfo{pages}{045001}.
\bibitem[{Gelfand and Fisher(1990)}]{Gelfand:1990vz}
\bibinfo{author}{M.~P. Gelfand}, \bibinfo{author}{M.~E. Fisher},
\newblock \bibinfo{title}{Finite-size effects in fluid interfaces},
\newblock \bibinfo{journal}{Physica A: Statistical Mechanics and its
  Applications} \bibinfo{volume}{166} (\bibinfo{year}{1990})
  \bibinfo{pages}{1--74}.
\bibitem[{Lapujoulade(1994)}]{Lapujoulade:1994uw}
\bibinfo{author}{J.~Lapujoulade},
\newblock \bibinfo{title}{The roughening of metal surfaces},
\newblock \bibinfo{journal}{Surface Science Reports} \bibinfo{volume}{20}
  (\bibinfo{year}{1994}) \bibinfo{pages}{195--249}.
\bibitem[{Saito(1996)}]{Saito:1996vo}
\bibinfo{author}{Y.~Saito}, \bibinfo{title}{Statistical Physics of Crystal
  Growth}, \bibinfo{publisher}{World Scientific, Singapore},
  \bibinfo{year}{1996}. \URLprefix \url{https://doi.org/10.1142/3261}.
  \DOIprefix\doi{doi:10.1142/3261}.
\bibitem[{Mendelev and Srolovitz(2002)}]{Mendelev:2002wv}
\bibinfo{author}{M.~I. Mendelev}, \bibinfo{author}{D.~J. Srolovitz},
\newblock \bibinfo{title}{Domain wall migration in 3-d in the presence of
  diffusing impurities},
\newblock \bibinfo{journal}{Interface Science} \bibinfo{volume}{10}
  (\bibinfo{year}{2002}) \bibinfo{pages}{243--250}.
\bibitem[{Gibbs(1948)}]{Gibbs}
\bibinfo{author}{J.~W. Gibbs}, \bibinfo{title}{The collected works of J. W.
  Gibbs}, volume~\bibinfo{volume}{1}, \bibinfo{publisher}{Yale University
  Press}, \bibinfo{address}{New Haven}, \bibinfo{year}{1948}.
\bibitem[{Frolov and Mishin(2015)}]{Frolov:2015ab}
\bibinfo{author}{T.~Frolov}, \bibinfo{author}{Y.~Mishin},
\newblock \bibinfo{title}{Phases, phase equilibria, and phase rules in
  low-dimensional systems},
\newblock \bibinfo{journal}{J. Chem. Phys.} \bibinfo{volume}{143}
  (\bibinfo{year}{2015}) \bibinfo{pages}{044706}.

\end{thebibliography}

\newpage{}

\begin{table}[ht]
\centering \caption{Table of physical and normalized variables in this model.}
\bigskip{}

\begin{tabular}{lcc}
\hline 
Variable & Physical & Normalized\tabularnewline
\hline 
\hline 
System dimensions & $Xa,Ya$ & $X,Y$\tabularnewline
Temperature & $T$ & $k_{B}T/J_{gg}$\tabularnewline
Total energy & $E$ & $E/J_{gg}$\tabularnewline
Heat capacity & $C$ & $C/k_{B}$\tabularnewline
Solute--interface interaction energy & $J_{sg}$ & $J_{sg}/J_{gg}$\tabularnewline
Solutes interaction energy & $J_{ss}$ & $J_{ss}/J_{gg}$\tabularnewline
Time & $t$ & $t\nu_{0}$\tabularnewline
GB energy & $\gamma$ & $\gamma a/J_{gg}$\tabularnewline
GB stiffness & $\alpha$ & $\alpha a/J_{gg}$\tabularnewline
GB length & $L$ & $L/(Xa)$\tabularnewline
Wave number & $k_{n}$ & $k_{n}a$\tabularnewline
Fourier amplitude & $A_{n}$ & $A_{n}/a$\tabularnewline
Mean squared GB width & $w^{2}$ & $w^{2}/a^{2}$\tabularnewline
Intrinsic GB width & $w_{I}$ & $w_{I}/a$\tabularnewline
Chemical potential & $\mu$ & $\mu/J_{gg}$\tabularnewline
\hline 
\end{tabular}\label{tab:variables}
\end{table}

\newpage\clearpage{}
\begin{center}
\setcounter{figure}{0}\setcounter{equation}{0}\setcounter{page}{1}
\par\end{center}

\begin{center}
\textbf{\large{}SUPPLEMENTARY INFORMATION}{\large\par}
\par\end{center}

\begin{center}
\textbf{\large{}A model of thermodynamic stabilization of nanocrystalline
grain boundaries in alloy systems}{\large\par}
\par\end{center}

\begin{center}
{\large{}Omar Hussein and Yuri Mishin\bigskip{}
}{\large\par}
\par\end{center}

\lyxaddress{\begin{center}
{\large{}Department of Physics and Astronomy, MSN 3F3, George Mason
University, Fairfax, Virginia 22030, USA}
\par\end{center}}

\section*{Grain boundary free energy in alloys from thermodynamic integration}

In this supplementary file, we provide a more detailed description
of the thermodynamic integration procedure applied in section 6 of
the main text.

\begin{figure}[!htb]
\centering \includegraphics[width=1\linewidth]{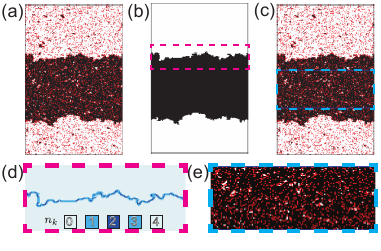} \caption{Example of GB shape analysis for thermodynamic integration of the
GB free energy in alloys. (a) Initial structure containing a fuzzy
GB and background noise in the grains. (b) ``Clean'' structure after
removing the noise in the grains. The dashed red lines outline a box
containing a GB. (c) The box bounded by the dashed blue lines represents
GB interiors and is used for computing the solute concentration in
the grains ($c_{g}$). This box is shown separately in (e). (d) The
GB portion inside the dashed red box in (b). The GB cells are color-coded
by the values of the parameter $n_{k}$ (the tracker of GB cells).
Such GB structures were used to compute the GB length $L$.}
\label{fig:snapshots}
\end{figure}

The grain boundary (GB) free energy in alloys was calculated in a
periodic system comprising two parallel GBs (Fig.~\ref{fig:snapshots}).
The calculation is based on the two-dimensional form of the Gibbs
adsorption equation
\begin{equation}
d(\gamma L)=-\tilde{N}d\mu,\quad\text{}T=\text{const}.\label{eqn:adsorption}
\end{equation}
Here, $\gamma$ is the GB free energy, $L$ is the GB length, $\mu$
is the chemical potential of the solute atoms, and $\tilde{N}$ is
the total amount of solute segregation per GB. The latter is defined
by the equation: 
\begin{equation}
\tilde{N}=(c-c_{g})\frac{N}{2},\label{eqn:GB_seg}
\end{equation}
where $c_{g}$ is the solute concentration in grain regions unaffected
by the GBs, $N$ is the number of cells in the system, and $c$ is
the average solute concentration in the alloy. The factor of two accounts
for the presence of two GBs.

By integrating Eq.(\ref{eqn:adsorption}), we derive the following
expression for the GB energy as a function of temperature and chemical
potential: 
\begin{equation}
\gamma(T,\mu)=\frac{1}{L(T,\mu)}\left[\gamma(T,\mu_{*})L(T,\mu_{*})-\int_{\mu_{*}}^{\mu}\tilde{N}(T,\mu^{\prime})d\mu^{\prime}\right].\label{eqn:Alloys_GB_energy}
\end{equation}
The asterisk denotes the chemical potential in the reference state
$(T,\mu_{*})$. The ensemble-averaged GB length $L$ can differ from
the system size $X$ due to capillary fluctuations. The reference
chemical potential $\mu_{*}$ is chosen to achieve the smallest possible
solute concentration at which $\tilde{N}$ can still be reliably extracted
from the simulations. The solute-free $\gamma(T)$ values were calculated
in section 4.2. Metropolis Monte Carlo simulations were performed
on a $128\times256$ system with temperatures between 0.3 and 2.0.
To ensure that the system has reached thermodynamic equilibrium and
collect sufficient statistics, the simulations comprised about $10^{9}$
Monte Carlo steps. 

The integration in Eq.~(\ref{eqn:Alloys_GB_energy}) requires several
ingredients, which were obtained as follows. First, we extracted the
GB shapes from the simulation snapshots. The challenge in doing so
is caused by the fuzziness of the GBs and the background noise in
the grains. The noise in the gains is caused by the numerous cells
of opposite color, either isolated or aggregated into small clusters.
This noise is especially significant at high temperatures. An example
is shown in Fig.~\ref{fig:snapshots}(a). An algorithm was developed
to identify the ``wrong'' color cells/clusters lying inside the grains
and reverse their color to that of the grain (Fig.~\ref{fig:snapshots}(b)).
This grain-cleaning procedure was applied to all snapshots saved during
the simulations. Next, the GBs were identified and the instantaneous
average GB positions $\bar{y}$ were calculated:
\begin{equation}
\bar{y}=\frac{1}{X}\sum_{n=1}^{X}y_{n},\label{eqn:GB_position}
\end{equation}
where $y_{n}$ are the $y$-coordinates of the GB cells (the $y$-axis
is normal to the GBs). The instantaneous positions of the two GBs
were monitored during the simulations, allowing us to choose regions
containing the GBs and regions located entirely inside the grains.
For example, Fig.~\ref{fig:snapshots}(b,c) shows such regions outlined
by the dashed red and blue lines, respectively. The grain regions
were used to compute the solute concentration in the grains, $c_{g}$,
required for computing $\tilde{N}$ by Eq.~(\ref{eqn:GB_seg}). Another
required quantity is the average solute concentration in the system,
$c$, which was found by simply counting the solute atoms.

The next ingredient, the GB length $L$, was calculated as follows:
\begin{equation}
L=\left\langle \frac{1}{4}\sum_{k}\lambda_{k}\right\rangle ,\quad\text{where }\lambda_{k}=\begin{cases}
1 & \text{if }n_{k}\neq0,\\
0 & \text{if }n_{k}=0.
\end{cases}\label{eqn:GB_length}
\end{equation}
Here, the angular brackets indicate ensemble averaging, and the factor
of $1/4$ eliminates the double-counting of pairs of GB cells and
accounts for the presence of two GBs in the system. As an example,
Fig.~\ref{fig:snapshots}(d) shows a zoom-in view of one of the GBs
with GB cells color-coded by the variable $n_{k}$.

\begin{figure}[!htb]
\centering \includegraphics[width=0.37\linewidth]{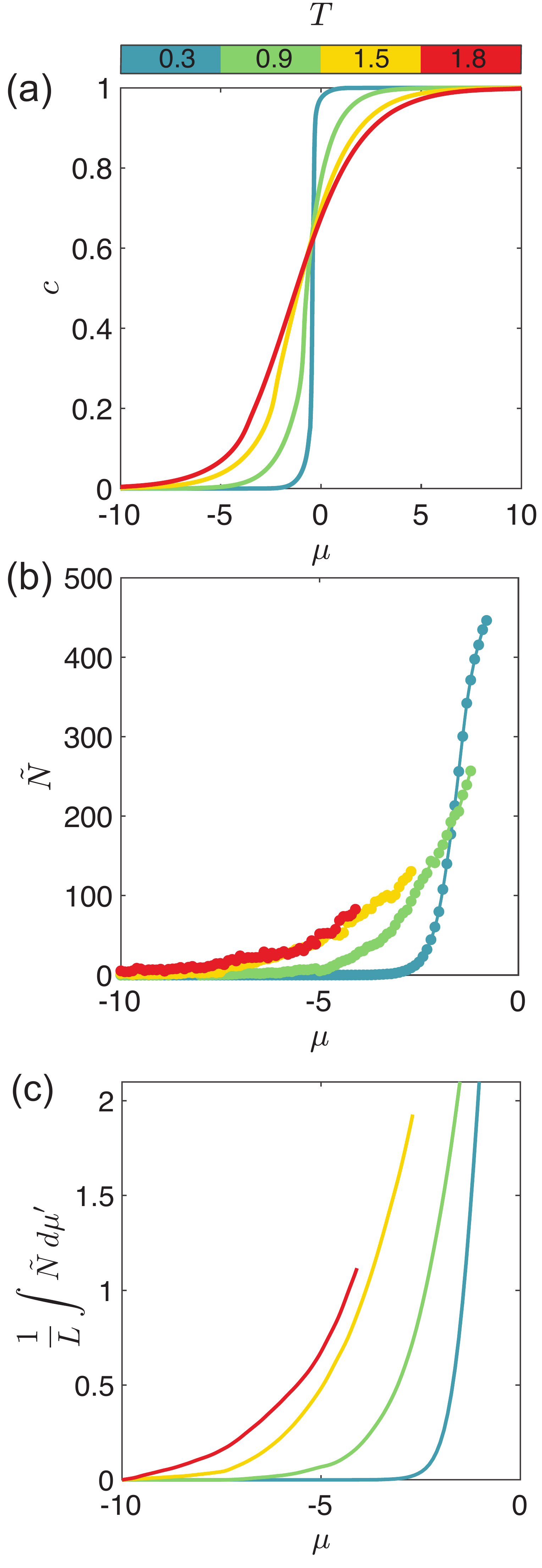} \caption{Steps of the GB free energy calculation in alloys at the temperatures
of 0.3, 0.9, 1.5, and 1.8. by thermodynamic integration. (a) The relationship
between the alloy concentration $c$ and the chemical potential $\mu$.
(b) The total amount of GB segregation $\tilde{N}$ as a function
of chemical potential $\mu$. The data points represent individual
simulations. The curves were obtained by quadratic spline interpolation.
Each curve starts and ends at critical points where the bicrystalline
structure disintegrates. (c) The integral in Eq.(\ref{eqn:Alloys_GB_energy})
as a function of $\mu$. The integration was performed using the quadratic
spline interpolation of the function $\tilde{N}(\mu)$.\label{fig:Supplem_plot-2}}
\label{fig:plots}
\end{figure}

The described calculations of $c$, $\tilde{N}$, and $L$ were repeated
for a set of chemical potentials and temperatures. Fig.~\ref{fig:Supplem_plot-2}(a,b)
shows $c$ and $\tilde{N}$ as a function of $\mu$ for four representative
temperatures. The reference chemical potential was chosen to be $\mu_{*}=-10$.
The numerical integration in Eq.(\ref{eqn:Alloys_GB_energy}) was
performed using a quadratic spline interpolation of $\tilde{N}$ as
a function of $\mu$, resulting in continuous curves shown in Fig.~\ref{fig:Supplem_plot-2}(c).
While Eq.(\ref{eqn:Alloys_GB_energy}) yields the function $\gamma(T,\mu)$,
the more convenient function $\gamma(T,c)$ was obtained from the
plots of $c$ versus $\mu$ exemplified in Fig.~\ref{fig:Supplem_plot-2}(a).

\end{document}